\documentclass[a4paper,11pt]{article}
\usepackage{booktabs}
\usepackage{bm}
\usepackage{float}
\usepackage{latexsym}
\usepackage{dcolumn}
\usepackage{amsfonts,amssymb}
\usepackage{graphicx}
\usepackage{epsfig}
\usepackage{psfrag}
\usepackage{nccmath}
\usepackage{moresize}
\usepackage{enumerate}
\usepackage[hang,flushmargin]{footmisc}
\usepackage[titletoc,toc]{appendix}
\usepackage{lipsum}
\usepackage{subcaption}
\usepackage{hyperref}
\usepackage{rotating}
\usepackage{multirow}
\usepackage{multicol}
\usepackage{xfrac}
\usepackage{mathtools}
\usepackage{nccmath}
\usepackage{cite}
\usepackage{placeins}
\usepackage{tikz}
\usetikzlibrary{positioning}
\usepackage{textalpha}
\graphicspath{{figures/}} % Location of the graphics fileshttps://www.overleaf.com/project/608a9528e951a97ae206ea82
\usepackage[font=small,labelfont=bf]{caption}
\usepackage{wrapfig}
\usepackage{authblk}
\usepackage{cite}
\usepackage[utf8]{inputenc}
\usepackage[T1]{fontenc}
\usepackage{enumitem}
\usepackage{blindtext}

\title{\bf Generic modification of gravity, late time acceleration and 
Hubble tension}
\author[1]{ Mayukh R. Gangopadhyay\thanks{mayukh$\_$ccsp@sgtuniversity.org}}
\affil{\small \it  Centre for Cosmology and Science Popularization(CCSP), SGT University, Gurugram 122505, India}
\author[1]{ Shibesh K. Jas Pacif\thanks{shibesh$\_$ccsp@sgtuniversity.org}}
\author[1,2,3]{M.~Sami\thanks{samijamia@gmail.com;~sami$\_$ccsp@sgtuniversity.org}}
\affil[2]{\small \it Center for Theoretical Physics, Eurasian National University, Astana 010008, Kazakhstan}
\affil[3]{\small \it Chinese Academy of Sciences,52 Sanlihe Rd, Xicheng District, Beijing }
\author[1]{Mohit K. Sharma\thanks{mr.mohit254@gmail.com}}

\begin{document}
\maketitle
\date{}
\begin{abstract}
We consider a scenario of large-scale modification of gravity that does not invoke extra degrees of freedom but includes coupling between baryonic matter and dark matter in the Einstein frame. The total matter energy density follows the standard conservation, and evolution has the character of deceleration in this frame. The model exhibits interesting features in the Jordan frame realised by virtue of a disformal transformation where individual matter components adhere to standard conservation but gravity is modified. A generic parametrization of disformal transformation leaves thermal history intact and gives rise to late time acceleration in the Jordan frame, which necessarily includes phantom crossing, which, in the standard framework, can be realised using at least two scalar fields. This scenario is embodied by two distinguished features, namely, acceleration in the Jordan frame and deceleration in the Einstein frame, and the possibility of resolution of the Hubble tension thanks to the emergence of phantom phase at late times.
\end{abstract}

\section{Introduction}

The hot big-bang model has several remarkable successes to its credit, 
including prediction of the expanding universe, microwave background 
radiation, synthesis of light elements in the early universe, and growth of 
structure via gravitational instability. 
The model, however, suffers from inbuilt inconsistencies related to early times$-$ for instance, flatness problem, horizon problem and late stages of evolution$-$ age puzzle. Because the matter-dominated era contributes the most to the age of Universe, the late time slow down of Hubble expansion must be invoked, allowing the Universe to spend more time before reaching $H_0$ and thus improving the age of Universe. The only way to accomplish the Hubble slowdown at late stages of evolution is to introduce a late time acceleration \cite{DEbook,Copeland,Sami,Sami2,Sami-notes,Trodden,Sahni,Frieman,Caldwell,
Peri,Frieman2,Carroll,Pady,age,LopezCorredoira2017,age2,Damjanov,age3,LopezCorredoira2018}.

The inconsistencies of the hot big bang are successfully resolved by complementing the model with early and late time phases of accelerated expansion. Inflation not only addresses the early time inconsistencies of the model but also provides a mechanism for the generation of primordial density perturbations responsible for structure in the universe \cite{linde,Kolb,mrg1,mar1,sharma,Bhattacharya,Bhattacharya:2020zap1}. Inflation can be achieved using scalar field(s) or gravity modification at small scales, as in the Starobinsky model \cite{starobinsky}.
As for the late time cosmic acceleration, it may either be caused by an exotic fluid of large negative pressure dubbed dark energy (quintessence) or by a large scale modification of gravity \cite{Sahni,Frieman,Caldwell,Peri,Frieman2,Carroll,Pady,Ratra,Wetterich,Ratra2,Steinhardt,ijmpds,SAMI-NOJ}. 
As mentioned before, late time acceleration is the only known remedy for the age puzzle in the hot big-bang scenario.
Age considerations, however, do not accurately constrain the equation of state 
of dark energy and its contribution to the total energy budget of the Universe;
necessary is done by supernovae Ia 
\cite{Riess,Perlmutter,jla,pantheon} 
and other indirect observations 
\cite{Seljak,BasilakosH0,reiss2019,reiss2021,SBF,trgb,birrer,chen,wong,huang,snII,TFR,hotens,Efstathiou2014,Hub19}.

Different schemes of large-scale modifications of gravity have been investigated in the literature \cite{antonio,derham,massive2,massive3,massive4,KHOURY2016,shibesh,jwang,kBamba,LPC1,vainsh1,vainsh2,chamel0,chamel1,chamel2,EXTENDED,CAPOZZI,sponts,ballardini1,ballardini3,ballardini2,ballardini4}. 
Most of these schemes reduce to Einstein's gravity plus extra degrees of freedom, which are non-minimally coupled. One extra scalar degree of freedom exists in $f(R)$ gravity; massive gravity has three extra degrees of freedom (two vector and one scalar) 
\cite{derham,massive2}.
The scalar degree of freedom to comply with the requirement of late time acceleration should be light with a mass of the order of $10^{-33}$ eV.
These extra degrees of freedom are directly coupled to matter through universal coupling similar to graviton, resulting in an effective doubling of the Newton constant $G$ in the case of $f(R)$ gravity, wreaking havoc locally where Einstein gravity complies with observation to one part in $10^5$ \cite{Copeland}.  
In massive gravity, only mass-less scalar degree of freedom survives in the decoupling limit relevant to local physics and is beautifully screened due to an inbuilt arrangement known as Vainstein mechanism \cite{vainsh1,vainsh2}. 
Unfortunately, massive gravity fails for other reasons \cite{massive3,massive4}. 
The chameleon mechanism is used in the $f(R)$
theory, where the extra degree of freedom becomes heavy locally and escapes dynamics, but remains light over large distances and causes late time acceleration 
\cite{LPC1,chamel0, chamel1}. 
To a surprise at the onset, it turns out that accurate local screening leaves no scope for late time acceleration in chameleon theories \cite{LPC1,chamel0, chamel1, chamel2,EXTENDED,CAPOZZI,sponts}. 
In fact, in this case, acceleration can not be distinguished from the one caused by cosmological constant or quintessence. It sounds quite strange that screening at the level of a solar system with a size of the order of $10^{14}$ cm influences physics at the horizon scale ($10^{28}$ cm). However, it might not be that strange if we think in terms of relative matter densities, namely, $10^{-24}$ gm/cc in the solar system versus the critical density, $10^{-29}$gm/cc, with a difference of five orders of magnitude; ironically, Einstein gravity is accurate to one part in $10^5$ in the solar system.
Thus, the purpose of large scale modification due to extra degrees of freedom invoked to mimic late time acceleration is grossly defeated. A generic large-scale modification of gravity should have the following distinguished property: {\it Acceleration in the Jordan frame and no acceleration in the Einstein frame}. In that case, acceleration can be attributed to the modification of gravity. 

In $f(R)$ theories, for instance, matter follows standard conservation in the 
Jordan frame, and gravity is modified $-$ its action differs from 
Einstein-Hilbert; in the Einstein frame, the Lagrangian is diagonalized, and 
gravity is standard, but there is a scalar field with direct coupling to 
matter. Acceleration in the Einstein frame is not removed by proper screening 
of the scalar degree of freedom, and $f(R) $ theory fails to meet the 
criterion of generic modification (see \cite{antonio}, and references therein). A novel scheme of large-scale modification 
of gravity was proposed in \cite{KHOURY2016} see also Ref.\cite{shibesh} 
where one assumes coupling between baryonic and dark matter in the Einstein 
frame such that baryonic+dark matter follows the standard conservation in the 
Einstein frame and there is no acceleration there. On the other hand, one can 
remove the said coupling by going to the Jordan frame via disformal 
transformation, where baryonic and dark matter individually follow standard 
conservation but gravity action is different from Einstein-Hilbert. In this 
case, disformal coupling can be parameterised to yield late time acceleration 
in the Jordan frame which should naturally be attributed to modification of 
gravity. Obviously, the criterion of generic modification is satisfied in 
necessarily manifests in this case in a generic way.  It was demonstrated in 
the \cite{shah} that the Hubble tension gets resolved in this case. It should
be noted that phantom phase naturally appears here and does not require the 
presence of a phantom field. In the model under consideration,  deviations 
from  $\Lambda$CDM take place only at late stages of evolution around the 
present epoch where phantom phase appears automatically. It should be noted 
that one needs at least two scalar fields to realise the phantom crossing 
which is naturally mimicked here by assuming coupling between baryonic and 
dark matter components in the Einstein frame.  

In this article, we briefly describe the aforesaid modification of gravity and Hubble tension between the Planck and local measurements of the Hubble parameter.  We  present details as how the tension gets 
naturally resolved in the framework under consideration.

%%%%%%%%%%%%%%%%%%%%%%%%%%%%%%%%%%%%%%%%%%%%%%%%%%%%%%%%
\section{Hubble Tension Overview}

The fact that the present-day cosmic expansion rate $H_0=67.36 \pm 0.54$\,km/s/Mpc 
\cite{Planck18}
predicted by the $\Lambda$CDM model from the cosmic microwave background (CMB) radiation observations by the Planck satellite does not agree with the low-redshift observations on the Hubble constant gives rise to the ``Hubble Tension'' problem \cite{reiss2021,RiessH02018,Riess2011,RiessH02016,RiessH02017,Melchiorri2016,Jarahetal2017,Feeney2018}. Specifically, the Supernovae $H_0$ for the Equation of State (SH0ES) collaboration estimates a substantially higher value of the Hubble constant i.e. 
$73.5 \pm 1.4$\,km/s/Mpc (which amounts to $4.2\,\sigma$ level discrepancy with Planck) using Cepheid calibrated supernovae Type Ia 
\cite{reiss2021}. 
Same level of discrepancies on $H_0$ have also been observed in 
various other low-$z$ measurements, including H0LiCOW's  $73.3^{+1.7}_{-1.8}$\,km/s/Mpc 
\cite{holicow} 
and Megamaser Cosmology Project's $73.9\pm3.0$\,km/s/Mpc  
\cite{megamaser,Reidetal2013,Kuoetal2013,Gaoetal2016} 
results.

Despite the observable disparity, several theoretical solutions have 
been proposed to deal with this problem. These solutions may be divided into three main types:
\begin{itemize}
    \item {\bf Early-time Modifications}: In cosmology, the positions of 
    the acoustic peaks in the CMB temperature and anisotropy spectra are 
    among the most accurately measured quantities 
    \cite{Planck18}. 
    These acoustic peaks 
    helps in determining the size of the sound horizon at the recombination epoch. In order to attempt to modify the sound horizon one needs to 
    introduce new physics during the pre-recombination epoch that 
    deform $H(z)$ at $z >1100$ 
    \cite{kam,kam2,Hub2,Hub3,Hub4,Hub5,Hub6,Hub7,Hub8,Hub9,Hub10,Hub11}. 
    One such modification is motivated from some string-axiverse-inspired scenarios for dark energy, in which 
    dark energy density at early times behaves like the cosmological 
    constant but then decays quickly. However, in this approach Hubble constant can at-most shift to $1.6$\,km/s/Mpc at redshift $z \simeq 1585$ 
    \cite{kam}. 
    Other approaches, such as modifying the standard model neutrino sector 
    \cite{trodden,Hub12}, 
    additional radiation 
    \cite{baren}, 
    primordial magnetic fields 
    \cite{pogosian}, 
    or adjusting basic constants with the goal of lowering the sound horizon at recombination 
    \cite{pogosian2}, 
    are insufficient to properly answer the Hubble Tension problem. 
    Also, they expect large growth of matter perturbations than reported by redshift space distortion (RSD) and weak lensing (WL) data, worsening the $\Omega_{0m}-\sigma_8$ tension 
    \cite{perivola}. 
    
    \item {\bf Late-time Modifications}: In this approach, one considers late-time alternative forms of DE 
    \cite{DEbook}, 
    unified dark fluid models 
    \cite{Copeland,Sami-notes}
    where the Dark Matter (DM) and DE behave as a single fluid, alternative gravitational theories including either modified versions of GR or new gravitational theories beyond GR 
    \cite{antonio}, 
    interacting DE models 
    \cite{DEbook,Copeland}
    in which DM and the DE interact with each other in a non-gravitational way \cite{Hub17,Melchiorri2017b}, (for more refs, please see \cite{PLB2017,ApJL2015,ApJ2017,EPL2018x3,ChenRatra2011,GWH0second}).  In the latter scenario, DE-DM 
    interaction provide a possible solution to the cosmic coincidence problem, and also can explain the phantom DE regime without any scalar field having a negative kinetic term. It is argued that some of these 
    models also not able to fully resolve the Hubble Tension problem 
    \cite{Benevento}. 
    
    \item {\bf Late-time transition of SnIa absolute magnitude $M$}: 
    The shifting of $M$ to a lower value by $-0.2$ at redshift $z_t \simeq 0.1$ is also a possible approach to address the Hubble Tension problem. Such a reduction in $M$ at $z > z_t$ may be caused, for example, by a comparable transition of the effective gravitational constant $G_{eff}$, which would result in a rise in the SnIa intrinsic brightness at $z > z_t$. This class of models has the potential to entirely solve the Hubble Tension problem while also addressing the growth tension by slowing the growth rate of matter perturbations 
    \cite{Hub16,TammannReindl2013,JangLee2017}. 
\end{itemize}

In this paper, we will show a proposition to solve the Hubble Tension issue which relies entirely on dark matter and the baryonic matter.
The mechanism depends on the coupling of dark matter and baryonic matter via an effective metric. We will demonstrate that this model may well fit cosmological observables under the assumption of a set of parametrization of the effective metric.

\begin{figure}[htb!]
    \centering
    \includegraphics[scale=0.6]{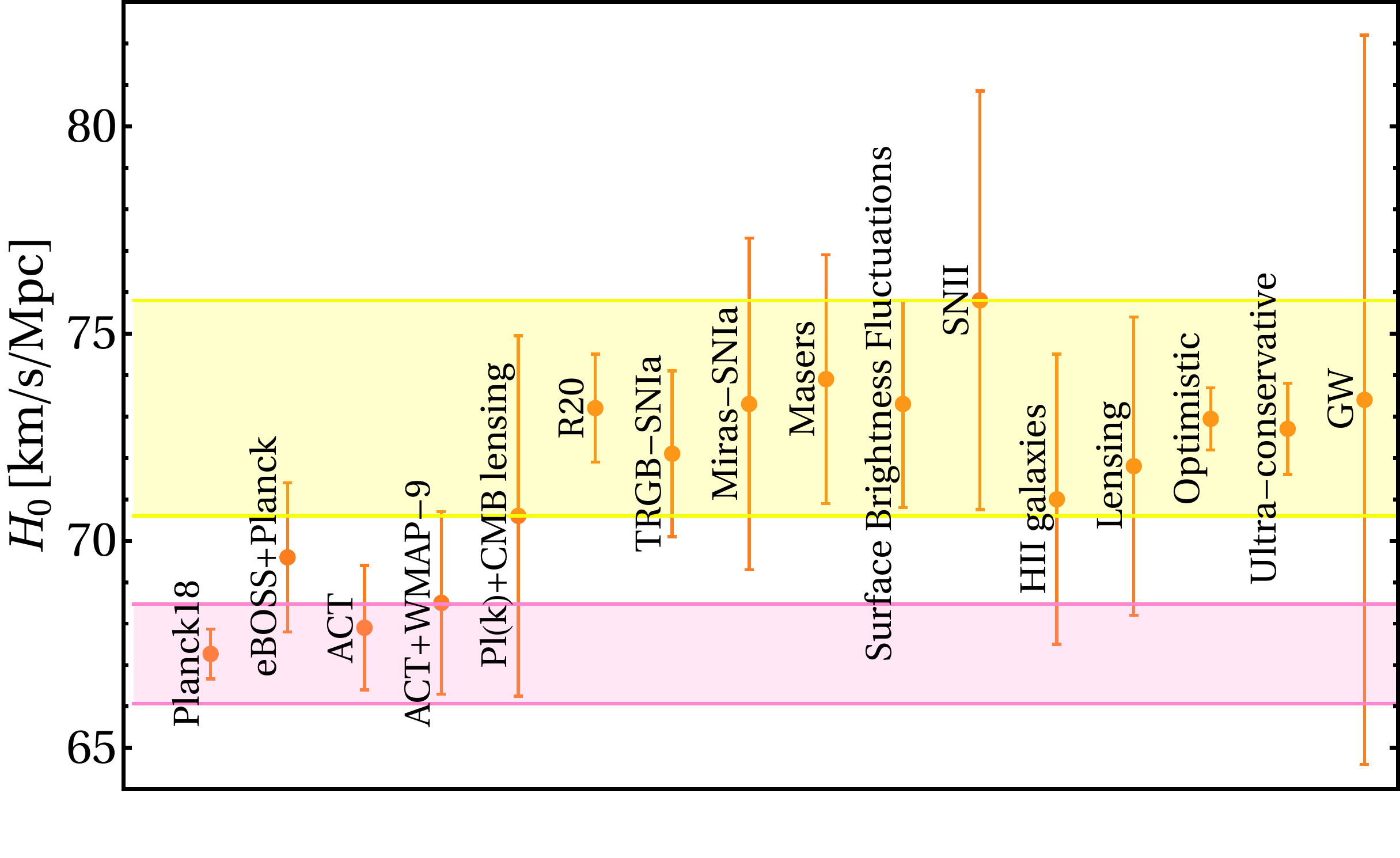}
    \caption{Whisker plot(following \cite{hotens}) with $68\%$ confidence 
    limit constraints of the Hubble constant $H_0$ through several direct
    and indirect measurements. The yellow horizontal band corresponds to
    the $H_0$ value from $SH0ES$ Team ($H_0 = 73.2 \pm 1.3$ km/s/Mpc at $68\%$ confidence limit)
    \cite{reiss2021} 
    and the pink vertical band corresponds to the $H_0$ value as 
    reported by \textit{Planck 2018} 
    \cite{Planck18} 
    within a vanilla $\Lambda$CDM scenario.}
    \label{plot1}
\end{figure}

\section{Coupling between Baryonic and Dark matter components in the Einstein frame}
In what follows we shall discuss a scenario which assumes coupling between Baryonic and Dark matter in the Einstein frame such that the sum total of matter components adheres to standard conservation in this frame. One can then imagine to go back to the physical frame, namely, the Jordan frame via a disformal coupling where the matter components individually satisfy the standard conservation but Einstein-Hilbert action is modified. A suitable parametrization that conforms to $\Lambda$CDM in the past might give rise to late time acceleration which has the characteristic of
supper-acceleration dubbed phantom behaviour.

\subsubsection{Disformal coupling between matter components } 
\label{disf}
In this sub-section, we  discuss a framework which operates through a mechanism that assumes interaction between BM and DM in the Einstein frame.  The latter can be realized by assuming  the following action in the Einstein
frame\cite{KHOURY2016}, 
\begin{equation}
\label{lag1}
\mathcal{S}=\int d^4 x \left(\frac{1}{16\pi G}\sqrt{-g}\mathcal{R}+\mathcal{L}%
_{DM}\{g_{\mu \nu }\}+\mathcal{L}_{BM}\{\widetilde{g}_{\mu \nu }\}\right),
\end{equation}
where $\widetilde{g}_{\mu \nu }$
and  $g_{\mu \nu }$ denote the Jordan and Einstein frame metric, respectively. 
The action (\ref{lag1}) gives rise to coupling between DM and BM in the Einstein frame where evolution would have decelerating character. 
We should then construct
the Jordan frame metric,
$\widetilde{g}_{\mu \nu }$ from 
$g_{\mu \nu }$
 and parameters
that define the dark matter. To this effect, we shall make use of disformal transformation between the two frames. In the perfect fluid representation, the dark matter 
 then can be described by  a  scalar field, 
\begin{equation}
\mathcal{L}_{DM}=\sqrt{-g}P(X) ;~~ \quad X \equiv -g^{\mu \nu }\partial _{\mu }\Phi \partial _{\nu }\Phi  \, ,
\label{Ldm}
\end{equation}
By varying Eq.\,(\ref{Ldm}) with the Einstein frame metric ($g_{\mu\nu}$) we get the energy momentum tensor $T_{\mu\nu}$ of dark matter as,
\begin{equation}
T_{\mu \nu } \equiv \frac{2}{\sqrt{-g}}\frac{\delta \mathcal{S}_{DM}}{\delta g^{\mu\nu}} = 2P,_{X}\partial _{\mu }\Phi \partial _{\nu }\Phi +Pg_{\mu\nu } \, . 
\label{2a}
\end{equation}
where $\mathcal{S}_{DM}$ denotes the action for dark matter. The energy momentum tensor
can be cast in the  perfect fluid form,
\begin{equation} 
T_{\mu \nu
}=(\rho_{_{DM}}+P_{_{DM})}u_{\mu }u_{\nu }+P_{_{DM}}g_{\mu \nu } \,,
%&& \rho _{DM}=2P,_{X}(X)X-P(X) \, , \quad P_{DM}=P(X) \, , \quad u_{\mu }=-%
%\frac{1}{\sqrt{X}}\partial _{\mu }\Phi \, . 
\label{3}
\end{equation}
which when compared with the above expression (\ref{2a}), 
we get
\begin{eqnarray}
    P(X) &=& \rho_{_{DM}} \,,  \label{3a} \\ 
    2P_{,X}  \Phi_{,\mu} \Phi_{,\nu} &-& P(X) u_\mu u_\nu = \rho_{_{DM}} u_\mu u_\nu \,.
    \label{3b}
\end{eqnarray}
Moreover, by using the constraint: $g^{\mu\nu} u_\mu u_\nu = -1$, in 
Eq.\,(\ref{Ldm}), we get the following relation:
\begin{equation}
    u_\mu = -\frac{\Phi_{,\mu}}{\sqrt{X}} \,,
    \label{u}
\end{equation}
which determines that the DM fluid velocity components are sourced by the 
rate of change of DM field with the corresponding spacetime coordinates. 
By using Eq.\,(\ref{u}) back in (\ref{3b}), we get
\begin{equation}
\rho_{_{DM}} = 2 P_{{,X}} - P(X) \,.
\end{equation}

The energy momentum tensor for Baryonic matter  is given by,
\begin{equation}
\widetilde{T}_{BM}^{\mu \nu }\equiv \frac{2}{\sqrt{-\widetilde{g}}}\frac{\delta \mathcal{S}%
_{BM}}{\delta \widetilde{g}_{\mu \nu }} \,,
\label{6}
\end{equation}
where $\mathcal{S}_{BM}$ denotes the action for the Baryonic matter.
%It may be noted that the matter components are not coupled in the Jordan frame which implies that,
%\begin{eqnarray}
%&& \widetilde{\nabla}_{\mu }\widetilde{T}_{BM}^{\mu \nu }=0 \\
%&& \widetilde{\nabla}_{\mu }\widetilde{T}_{DM}^{\mu \nu }=0
%\label{6a1}
%\end{eqnarray}
%
%The
%Jordan frame metric $\widetilde{g}_{\mu \nu }$ can be expressed through 
%Einstein frame metric, $g_{\mu\nu}$, and  the dark matter field. We look for a  general relation between $\widetilde{g}_{\mu \nu}$, $g_{\mu \nu }$ and $\Phi$ dubbed disformal transformation,
%\begin{eqnarray}
%\widetilde{g}_{\mu \nu} =R^{2}(X)g_{\mu \nu }+S(X)\partial _{\mu }\Phi
%\partial _{\nu }\Phi ;~~S(X)\equiv \frac{R^{2}(X)-Q^{2}(X)}{X}\text{,}
%\label{4A}
%\end{eqnarray}%
%
%with $R$ and $Q$ are   arbitrary functions of X. 
%  Einstein equations are obtained by varying the action  with
%respect to Einstein frame metric,
%
%\begin{eqnarray}
%&& G_{\mu \nu }= 8\pi GT^{eff}_{\mu\nu} \, \\
%&&T^{eff}_{\mu\nu}= T^{DM}_{\mu \nu }+QR^{3}\widetilde{T}_{BM}^{k\lambda %}\left(
%R^{2}g_{k\mu }g_{\lambda \nu }+(2RR,_{X}g_{k\lambda }+S,_{X}\partial
%_{k}\Phi \partial _{\lambda }\Phi )\partial _{\mu }\Phi \partial _{\nu
%}\Phi \right)  , 
%\label{7}
%\end{eqnarray}%
%which tells us that in the absence of coupling, i.e., $Q=R=1$, the energy momentum tensor reduces to the sum of energy momentum tensors of the individual components. It should be noted that $T^{eff}_{\mu\nu}$ is conserved and no exotic behavior is expected in the Einstein frame.

\subsection{Equations of motion for DM field}

Since the Baryonic matter and DM follow geodesics corresponds to $\widetilde{g}_{\mu \nu}$ and $g_{\mu \nu}$, respectively, a disformal 
relation between these metrics, will eventually be translated to the 
coupling between these two forms of matter. Hence, their equations of motion 
in the Einstein frame will now be dependent on each other. 
It is also worth noting that in the Jordan frame, the matter components are not connected to one other; while the dynamics may appear cumbersome, the energy momentum tensors of each components are preserved individually, i.e.
\begin{equation}
    \widetilde{\nabla}_{\mu }\widetilde{T}_{BM}^{\mu \nu }=0 \,, \qquad 
    \widetilde{\nabla}_{\mu }\widetilde{T}_{DM}^{\mu \nu }=0 \,.
\end{equation}
Since it is a common perception that Einstein and Jordan frame metrics are 
related to each other by the conformal transformations but that has its 
own limitations when it comes to take into account the interaction(s) between 
the two. In particular, BM being pressureless results in vanishing sound 
speed, similarly, DM in the absence of any interaction with BM also gives 
rise to a vanishing sound speed. However, in presence of BM-DM 
interaction the DM sound speed can give rise to relativistic sound speed. 
As a consequence, DM will behave as a relativistic fluid which is 
undesirable as it can give rise to oscillatory perturbations. This problem 
is inevitable when the Jordan and Einstein frame metrics are conformally 
related to each other. Interestingly, this problem can be avoidable in 
presence of disformal coupling (see \cite{KHOURY2016}). 

A general form of the disformal coupling can be written as:
\begin{equation}
    \widetilde{g}_{\mu \nu} =R^{2}(X)g_{\mu \nu }+S(X)\Phi _{,\mu }
    \Phi _{,\nu } ;~~S(X)\equiv \frac{R^{2}(X)-Q^{2}(X)}{X} \,,
\label{disformal}
\end{equation}
with $Q$ and $R$ are arbitrary functions of $X$. Thus, it is evident that
\begin{equation}
    \sqrt{-\widetilde{g}} = Q R^3 \sqrt{-g} \,.
\end{equation}
Now, by varying the action (\ref{lag1}) with respect to $\Phi$, we get
\begin{equation}
    -\sqrt{-g} P_{_{,X}}\frac{\delta X}{\delta \Phi} + 
    P(X) \left(\frac{\delta \sqrt{-g}}{\delta \Phi} \right) + 
    \frac{\delta \mathcal{S}_{_{BM}}(\widetilde{g}_{\mu\nu})}{\delta \widetilde{g}_{\mu\nu}} \frac{\delta \widetilde{g}_{\mu\nu}}{\delta \Phi} = 0 \,,
    \label{action-var}
\end{equation}
where by using Eq.\,(\ref{disformal}), one can write
\begin{equation}
     \frac{\delta \mathcal{S}_{_{BM}}(\widetilde{g}_{\mu\nu})}{\delta \widetilde{g}_{\mu\nu}} \frac{\delta \widetilde{g}_{\mu\nu}}{\delta \Phi} = 
     \frac{QR^3\sqrt{-g}\widetilde{T}_{_{BM}}}{2} 
     \left[\frac{2S (\delta \Phi)_{,\mu} \Phi_{,\nu} }{\delta \Phi }  + 
     \Phi_{,\mu} \Phi_{,\nu} \left(\frac{\delta S}{\delta \Phi}\right) + 
     \frac{\delta (R^2 g_{\mu\nu})}{\delta \Phi}
     \right] \,.
     \label{deltaSBM}
\end{equation}
Since, 
\begin{eqnarray}
    S_{,\Phi} = S_{,X} X_{,\Phi} \, \quad \mbox{where} \quad X_{,\Phi} = -2 g^{\mu \nu} (\delta \Phi_{,\mu}) \Phi_{,\nu} \,,
\end{eqnarray}
one can re-express Eq.\,(\ref{deltaSBM}) as follows:
\begin{equation}
  \frac{\delta \mathcal{S}_{_{BM}}(\widetilde{g}_{\mu\nu})}{\delta \widetilde{g}_{\mu\nu}} \frac{\delta \widetilde{g}_{\mu\nu}}{\delta \Phi} = 
  \frac{QR^3\sqrt{-g}\widetilde{T}_{_{BM}}}{2}
  \left[ 
  \frac{2S (\delta \Phi)_{,\mu} \Phi_{,\nu} }{\delta \Phi }  - 
  \frac{2S_{_{,X}} g^{\alpha\beta} (\delta \Phi)_{,\alpha} \Phi_{,\beta} \Phi_{,\mu} 
  \Phi_{,\nu}}{\delta \Phi} + 
  \frac{2R g_{\mu \nu}\delta R }{\delta \Phi} + 
  \frac{R^2 \delta g_{\mu\nu}}{\delta \Phi}
  \right] .
\end{equation}
The first term in the r.h.s. of the above expression can be further expressed as
\begin{equation}
    \frac{QR^3\sqrt{-g}\widetilde{T}_{_{BM}}}{2}
    \left[ 
    2S (\delta \Phi)_{,\mu} \Phi_{,\nu} \right] = 
    \left( 
    \sqrt{-g}QR^3 S \widetilde{T}_{_{BM}}^{\mu\nu} \Phi_{,\nu} \delta \Phi 
    \right)_{,\mu} - 
    \left(\sqrt{-g}QR^3 S \widetilde{T}_{_{BM}}^{\mu\nu} \Phi_{,\nu}
    \right)_{,\mu} \delta \Phi \,,
    \label{first}
\end{equation}
similarly, the second term can be expressed as
\begin{eqnarray}
    2S_{_{,X}} g^{\alpha\beta} (\delta \Phi)_{,\alpha} \Phi_{,\beta} \Phi_{,\mu} \Phi_{,\nu} = 
    -\left( 
    QR^3 \sqrt{-g}\widetilde{T}_{_{BM}}^{\mu\nu} S_{_{,X}} g^{\alpha \beta}
    \Phi_{,\mu} \Phi_{,\nu} \delta \Phi
    \right)_{,\alpha} \\ + 
    \left(
    QR^3 \sqrt{-g}\widetilde{T}_{_{BM}}^{\mu\nu} S_{_{,X}} g^{\alpha \beta} 
    \Phi_{,\beta} \Phi_{,\mu} \Phi_{,\nu}
    \right)_{,\alpha} \delta \Phi \,,
    \label{second}
\end{eqnarray}
and the third term as
\begin{equation}
    \frac{QR^3 \sqrt{-g} \widetilde{T}_{_{BM}}^{\mu\nu}}{2} 
    \left(
    2Rg_{\mu\nu} R_{_{,X}} X_{_{\Phi}} 
    \right)= 
     QR^3 \sqrt{-g} \widetilde{T}_{_{BM}}^{\mu\nu} R_{_{,X}}^2 g_{\mu\nu}
    (g^{\mu\nu} \Phi_{,\nu})_{,\mu} \,.
    \label{third}
\end{equation}
Putting Eqs.\,(\ref{first}-\ref{third}) back in \ref{action-var}, we finally 
get the equations of motion of DM field in the Einstein frame i.e.
\begin{equation}
    \left[
    \left(
    2P_{_{,X}} + QR^3 \widetilde{T}_{_{BM}}^{\mu\nu} 
    \left(
    R_{_{,X}}^2 g_{\alpha\beta} + S_{_{,X}} \Phi_{,\alpha} \Phi_{,\beta}
    \right)
    g^{\mu \nu} - QR^3 S \widetilde{T}_{_{BM}}^{\mu\nu}
    \right) \sqrt{-g} \Phi_{,\mu}
    \right]_{,\nu} = 0 \,.
\end{equation}
Also note that variation of two metrices are related with each other as follows:
\begin{equation}
    \delta \widetilde{g}_{\mu \nu} = R^2 \delta g_{\mu \nu} + 
    \left[
    R^2 \delta g_{\mu\nu} + (2R R_{_{,X}} g_{\mu\nu} S_{_{,X}} \Phi_{,\mu} \Phi_{,\nu}) 
    \right]
    g^{\alpha \kappa} g^{\beta \lambda} \Phi_{,\alpha} \Phi_{,\beta} \delta g_{\kappa \lambda} \,,
\end{equation}
by using which one can write the Einstein equation as:
\begin{equation}
G_{\mu\nu} = 8 \pi G_{N} 
    \left[
    T_{\mu\nu} + Q R^3 \widetilde{T}^{\kappa \lambda}_{BM} 
        \left(
        R^2 g_{\kappa \mu} g_{\lambda \nu} + 
            \left(
            R^2_{_{,X}} g_{\kappa \lambda} + S_{_{,X}} \Phi_{,\kappa} \Phi_{,\lambda} 
            \right) \Phi_{,\mu} \Phi_{,\nu} 
        \right)
    \right] \,,
    \label{einstein}
\end{equation}
where $G_{\mu\nu}$ is the Einstein Tensor. This indicates that the energy momentum tensor reduces to the sum of the energy momentum tensors of the individual components in the absence of coupling, i.e., $Q=R=1$. 

\subsection{Dynamics in FRW Universe }
\label{sect:2}

For our analysis, we resort to the Einstein frame metric that 
satisfies the spatially homogeneous and isotropic background, given by
\begin{equation}
ds^{2}=-dt^{2}+a^{2}(t)\left( dx^{2}+dy^{2}+dz^{2}\right)  \label{1}
\end{equation}
due to which coupling functions $Q$ and $R$ depends only upon 
the scale factor.  Consequently, the Jordan frame metric is given by
\begin{equation}
\widetilde{g}_{\mu \nu }=diag\left(
-Q^{2}(a),R^{2}(a)a^{2},R^{2}(a)a^{2},R^{2}(a)a^{2}\right) \,.
\label{9}
\end{equation}
By using Eq.(\ref{einstein}), Friedmann and Raychaudhuri equations are 
respectively given as
\begin{eqnarray}
3H^{2}=8\pi G \rho_{{_T}} &\equiv& 
8\pi G_N\left[ \rho^{(eq)}_{_{DM}}\sqrt{\frac{X}{X^{(eq)}}}\left( \frac{%
a^{(eq)}}{a}\right) ^{3}-P+QR^{3}\widetilde{\rho}_{b}(a)\right] \,, \label{2} 
\\ 
2\frac{\ddot{a}}{a}+H^{2} &=& -8\pi G_N (P+P_{b})  \,, \label{4}
\end{eqnarray}
when superscript `eq' denotes the reference point at the matter-radiation equality epoch. Also, $P$($P_{b}$) designates the pressure of DM(BM) in the Einstein frame, such that $P_{b}\equiv QR^{3}\widetilde{P}_{b}$. Assuming 
both DM and BM to be pressure-less, it implies, 
$\widetilde{P}_{b}\simeq 0$ and $P\ll 2XP,_{X}\label{21}$). 

Due to the fact that BM obeys the conservation laws in the Jordan frame, 
one finds
\begin{equation}
\widetilde{\rho}_b(a) \simeq \frac{\rho_b^{(eq)}}{R^3}\left(\frac{ a^{(eq)}}{a} \right)^3  \,.
\end{equation}
Since the dynamics of the universe that is governed by $a(t)$ only depends 
on the pressureless BM in the Einstein frame (see Eq.\,(\ref{4})), it is easy 
to find that $a(t)\sim t^{2/3}$. From Eq.\,(\ref{2}), the total matter density 
in the Einstein can be expressed as
\begin{equation}
\rho_{_{T}}(a)=\left(\rho^{(eq)}_{_{DM}} \sqrt{\frac{X}{X_{(eq)}}}+Q(X)\rho^{(eq)}_{b}\right)\left(\frac{ a_{(eq)}}{a}\right)^3 \,,
\end{equation}
which implies that the quantity in the parenthesis  is constant for an arbitrary function $Q(X)$. It was demonstrated in Ref.\cite{KHOURY2016} that the system is plagued with  instability in case $Q=R$ (conformal coupling) and one should therefore focus on disformal transformation. For the sake of simplicity, we shall assume, $Q(a)\equiv 1$ (maximally disformal case). In that case we are left with one function $R$ to deal with.
In what follows, without the loss of generality, we shall adhere to the maximally disformal case, namely, $Q(a)\equiv 1$ leaving with a single  function $R$ to deal with.  Let us recall that we wish to have acceleration in the Jordan frame,
($\ddot {\widetilde{a}}>0$) and deceleration in the Einstein frame($\ddot {{a}}<0$ ).
The function, $R(a)$,  needs to be parametrized in a way that the thermal history is respected followed by accelerated expansion at late times, i.e., thereby the 
 physical scale factor, $\widetilde{a}=R(a) a$
 for the entire history and only at late stages of evolution it should 
 grows
sufficiently fast such that $\widetilde{a}$  experiences acceleration at late times. Indeed, 
assuming $R$ to be concave up, its growth at late times might compensate the effect of deceleration in $a(t)$
making $\ddot{\widetilde{a}}$ positive, 
\begin{equation}
   \ddot{\widetilde{a}}=\ddot{R}a+2\dot{R}\dot{a}+R\ddot{a} 
   \label{dc}
\end{equation}
where $\ddot{R}>0$ by assumption.
If $\dot{R}$ is large or $R$ increases fast at late times, it might compensate the last  term in (\ref{dc}) which has decelerating character.
In the sub-section to follow, we shall use convenient representations for the scale factor that would conform to the mentioned phenomenological consideration.
 %%%%%%%%%%%%%%%%%%
 
\subsection{ Polynomial parametrization} 
We now choose function $R(a)$ in accordance to the aforesaid requirement. We shall use the following Polynomial parametrization,
\begin{equation}
\label{m1}
a(\widetilde{a})=\widetilde{a}+\alpha \widetilde{a}^{2}+\beta \widetilde{a}^{3},
\end{equation}
where $\alpha$ and $\beta$ are constants. In respective frames, the scale factor and redshift have the following relationships: 
\begin{equation}
\widetilde{a}=\frac{\widetilde{a}_{0}}{1+\widetilde{z}}\text{ , \ \ }a=\frac{a_{0}}{1+z} \,,  
\label{5}
\end{equation}
Since we are working in spatially flat space-time, the physical scale factor can be normalized to one i.e. $\widetilde{a}_{0}=1$, as a result 
the Einstein frame scale factor gets rescaled by some factor. For instance,  $a_{0}=1+\alpha+\beta\neq 1$ (for $\alpha,\beta \neq 0$).

By using (\ref{m1}) and (\ref{5}), one may express the Hubble parameter in 
terms of $\widetilde{z}$ in the Jordan frame as
\begin{equation}
\widetilde{H}(\widetilde{z})=\widetilde{H}_{0} F_{Pl}(\alpha,\beta,\tilde{z})\equiv \widetilde{H}_{0}\frac{(1+\alpha +\beta )^{\frac{1}{2}%
}(1+2\alpha +3\beta )\left( 1+\widetilde{z}\right) ^{\frac{9}{2}}}{\left( \left(
1+\widetilde{z}\right) ^{2}+\alpha \left( 1+\widetilde{z}\right) +\beta \right) ^{%
\frac{1}{2}}\left( \left( 1+\widetilde{z}\right) ^{2}+2\alpha \left( 1+\widetilde{z}%
\right) +3\beta \right) } \,, 
\label{a3}
\end{equation}%
which can be used to obtain the  effective equation of state parameter,
\begin{equation}
\widetilde{w}_{eff}(\widetilde{z}) = -\frac{2\dot{\widetilde{H}}}{~3\widetilde{H}^2} = \frac{\alpha \left( 5+6\alpha +5\widetilde{z}\right)
(1+\widetilde{z})^{2}+\beta (14+23\alpha +14\widetilde{z})(1+\widetilde{z})+18\beta ^{2}%
}{3\{(1+\widetilde{z})^{2}+\alpha (1+\widetilde{z})+\beta \}\{(1+\widetilde{z}%
)^{2}+2\alpha (1+\widetilde{z})+3\beta \}} \,.  \label{a5}
\end{equation}
In order to extract the dark energy (DE) equation of state $\widetilde{w}_{de}$, from (\ref{a5}), we need to define the dimensionless fractional density parameters for cold matter and dark energy, $\Omega^{(0)}_{M eff}$ $\&$ $\Omega^{(0)}_{DE}$. This can be accomplished
%%%%%%%%
by expressing (\ref{a3}) in the standard form by isolating   
 the term proportional to $(1+\widetilde{z})^{3}$ i.e.

\begin{equation}
\frac{\widetilde{H}^{2}}{\widetilde{H}_{0}^{2}}=A(\alpha ,\beta )(1+\widetilde{z}%
)^{3}+A(\alpha ,\beta )f(\widetilde{z})\text{,}  \label{34}
\end{equation}%
where  
\begin{eqnarray}
A(\alpha ,\beta ) &=&(1+\alpha +\beta )(1+2\alpha +3\beta )^{2}\text{,}
\label{34A} \\
f(\widetilde{z}) &=&-5(1+\widetilde{z}^{2})\alpha -\alpha \left( 49\alpha
^{2}-48\beta \right) +(1+\widetilde{z})\left( 17\alpha ^{2}-7\beta \right) 
\notag \\
&&+\frac{(1+\widetilde{z})\alpha ^{6}-5(1+\widetilde{z})\alpha ^{4}\beta +\alpha
^{5}\beta +6(1+\widetilde{z})\alpha ^{2}\beta ^{2}-4\alpha ^{3}\beta ^{2}-(1+%
\widetilde{z})\beta ^{3}+36\alpha \beta ^{3}}{\left( \alpha ^{2}-4\beta \right)
\left( (1+\widetilde{z})^{2}+(1+\widetilde{z})\alpha +\beta \right) }  \notag \\
&&+\frac{%
\begin{array}{c}
128(1+\widetilde{z})\alpha ^{6}-64\alpha ^{7}-720(1+\widetilde{z})\alpha ^{4}\beta
+576\alpha ^{5}\beta +864(1+\widetilde{z})\alpha ^{2}\beta ^{2} \\ 
-1512\alpha ^{3}\beta ^{2}-135(1+\widetilde{z})\beta ^{3}+918\alpha \beta ^{3}%
\end{array}%
}{\left( \alpha ^{2}-4\beta \right) \left( (1+\widetilde{z})^{2}+2\alpha (1+%
\widetilde{z})+3\beta \right) }  \notag \\
&&+\frac{%
\begin{array}{c}
128(1+\widetilde{z})\alpha ^{8}-960(1+\widetilde{z})\alpha ^{6}\beta +192\alpha
^{7}\beta +2160(1+\widetilde{z})\alpha ^{4}\beta ^{2}-1296\alpha ^{5}\beta ^{2}
\\ 
-1512(1+\widetilde{z})\alpha ^{2}\beta ^{3}+2376\alpha ^{3}\beta ^{3}+162(1+%
\widetilde{z})\beta ^{4}-1053\alpha \beta ^{4}%
\end{array}%
}{\left( \alpha ^{2}-4\beta \right) \left( (1+\widetilde{z})^{2}+2\alpha (1+%
\widetilde{z})+3\beta \right) ^{2}} \,.  
\label{35}
\end{eqnarray}%
Casting the Friedmann equation in Jordan frame in terms of fractional energy density parameters, we have,
\begin{equation}
{\widetilde{H}^{2}}=\widetilde{H}_{0}^{2}\left[ \Omega _{Meff}^{\left( 0\right) }(1+%
\widetilde{z})^{3}+\Omega _{DE}^{\left( 0\right) }F(\widetilde{z})\right] \text{,}
\label{36}
\end{equation}%
where, $\Omega _{Meff}^{\left( 0\right) }\equiv A$, $\Omega
_{DE}^{\left( 0\right) }\equiv Af(0)$, and $F(\widetilde{z})\equiv {f(\widetilde{z})}/{f(0)}$. 
%{\bf pl do the same exercise for the second parametrization corresponding to (21), expanding exponential in the denominator  in Teller series is justified.} 
%%%%%%%%
By using Eq.\,(\ref{36}), the equation of state parameter for (effective) 
dark energy is then obtained in the Jordan frame as
\begin{eqnarray} 
\widetilde{w}_{eff}(\widetilde{z}) = \widetilde{w}_{M}\Omega^{(0)}_{M eff} + \widetilde{w}_{de}(\widetilde{z}) \, \Omega^{(0)}_{DE} \,, \\ 
\Longrightarrow ~~ \widetilde{\omega}_{de}(z)=\widetilde{\omega}_{eff}(\tilde{z})/\Omega^{(0)}_{DE} \qquad \mbox{where} \qquad \widetilde{w}_{M}=0  \,,
\label{wtotal}
\end{eqnarray}
where $\Omega^{(0)}_{M eff}$ and $\Omega^{(0)}_{DE}$ are the effective 
matter (dust-like) and DE density parameters, respectively. It is also important to mention that for $\alpha=-0.1523$ and $\beta=-0.0407$, the equation of state parameter for this parametrization approaches  its $\Lambda$CDM limit i.e., $w_{de}\to -1$ at the present epoch.

Let us take notice of the fact that the dark energy equation of state parameter might 
at some point take on a super negative value ($<-1$) with a general 
behaviour reflected by phantom crossing. It is important to note that a phenomena like this cannot be replicated by a quintessence field; at least two scalar fields are required to simulate the phantom crossing. 
It is intriguing that the aforementioned behaviour may results from the presence of disformal coupling between DM and BM. It should also be 
emphasized that the Einstein-Hilbert action gets modified at the expense of 
decoupling of DM and BM in the Jordan frame. However, the modification 
of gravity under consideration does not result in any more degrees of 
freedom it can still enable the realisation of acceleration at later times. Last but not least, acceleration in this framework is generically caused by modification of gravity $\hat{\rm a}$  {\it  la} {\it acceleration in Jordan frame and deceleration in Einstein frame.}

\subsection{ Exponential parametrization}

Let us now consider the exponential parametrization that is given below:
\begin{equation}
    a(\widetilde{a})=\widetilde{a}e^{\alpha \widetilde{a}} \,,
    \label{m2}
\end{equation}
such that the Hubble parameter and effective equation of state parameter 
are given by,
\begin{eqnarray} 
\label{b3}
&& \widetilde{H}(\widetilde{z})= \frac{\widetilde{H}_0 (1+\alpha)(1+\widetilde{z})^{5/2}}{1+\widetilde{z}+\alpha}\exp{\left(\frac{3\widetilde{z}\alpha}{2(1+\widetilde{z})}\right)}\equiv H_0 F_{Ex}(\tilde{z},\alpha) \,, \\
&& \widetilde{w}_{eff}(\widetilde{z})=\frac{5\alpha (1+\widetilde{z})+3\alpha ^{2}}{3(1+%
\widetilde{z})\left[ (1+\widetilde{z})+\alpha \right] }\text{ .}  
\label{b5}
\end{eqnarray}
%%%%%%%%%%%
Here, it should be noted that, in contrast to the situation in (\ref{m1}), 
the Friedman equation derived for the exponential parametrization (\ref{m2}) is not given in a standard form that is useful for carrying out the 
parametric estimation. In order to do this, we cast (\ref{b3}) in the form corresponding to (\ref{34}), for which we apply the following ansatz: 
\begin{eqnarray} \label{p1-standard}
\left(\frac{\widetilde{H}(\widetilde{z})} {\widetilde{H} _0}\right)^2 &=& (1+\alpha)^A (1+\widetilde{z})^3 + B\, e^{C \, \widetilde{z}} ,\\
\widetilde{H}(\widetilde{z}) & = & \widetilde{H} _0 F_{(exp)}(\tilde{z},\alpha),
\end{eqnarray}
where $A$, $B$ and $C$ are constants. It is then natural to identify $(1+\alpha)^A$ with the effective matter density i.e. $\Omega^{(0)}_{M eff}=(1+\alpha)^A$ and $B$ with the effective dark energy density $B=\Omega^{(0)}_{DE}$ parameters at the present epoch. The condition $\Omega^{(0)}_{M eff} + \Omega^{(0)}_{DE}=1$ yields a constraint on constants, namely, $(1+\alpha)^A+B=1$ leaving us with two unknowns, say, $A$ and $C$ 
to fit $F_{exp}(\tilde{z},\alpha)$ with (\ref{p1-standard}). 

Let us now implement the fitting in such a way that $\widetilde{H}(\widetilde{z}) = \widetilde{H_0}$ at the present epoch. 
Using  non linear model fitting, we find: $A= 3.4185$ and $C = 0.2896$. 
It is then straight forward to express the equation of state parameter 
for dark energy ($\widetilde{w}_{de}(\widetilde{z})$) as
\begin{equation}
\widetilde{w}_{de}(\widetilde{z}) = -\frac{\alpha  \, e^{-0.2896 \widetilde{z}} \left[\alpha +1.6667(\widetilde{z}+1)\right]}{\left[(\alpha +1)^{3.4185}- 1\right] (\widetilde{z}+1) (\alpha +\widetilde{z}+1)} \, ,
\end{equation}
where the fitted values of $B$ and $C$ have been used for $\Omega^{(0)}_{DE}$.  Let us note that the exponential parametrization mimics $\Lambda$CDM 
for $\alpha = -0.3896$,
%%%%%%%%%%%%%%%

\section{Observational datasets}
\label{method}

In this section, we demonstrate the parametric estimations for both \ref{m1} and \ref{m2} parametrizations from the late-time background level observational data. Three sets of data were specifically used in the 
analysis: the distance modulus measurements of type Ia supernovae (SNIa), observational Hubble data (OHD), and angular diameter distances obtained 
with water megamasers. A brief description of these datasets are given 
as below:

\paragraph{\textbf{Pantheon+MCT SnIa data}:}

The Hubble rate data points, i.e., $E(z_i) = H(z_i)/H_0$, reported in \cite{riess} for six distinct redshifts in the range of $z \in [0.07,1.5]$, are used in this study which efficiently compress the information of SnIa at $z<1.5$ that are utilized in the Pantheon compilation, and the 15 SnIa at $z >1$ of the CLASH Multy-Cycle Treasury (MCT) 
and CANDELS programs provided by the HST. The raw SnIa data were transformed into $E(z)$ \cite{riess,amendola} by parametrizing $E^{-1}(z)$ at those six redshifts. 
The dimensionless Hubble rate $\tilde{h}$ is defined as $\frac{\widetilde{H}(\widetilde{z})}{\widetilde{H_0}}$ and hence $\chi^2$ for the Supernova data is calculated as
\begin{equation}
\chi^2_{\rm{SN}}=\sum_{i,j}\left(E_i-\tilde{h}_i\right)\cdot c_{ij}^{-1}\cdot \left(E_i-\tilde{h}_i\right) \, ,
\end{equation}
where $c_{ij}$ is the correlation matrix between the data points.

\paragraph{\textbf{Observational Hubble Data (OHD)}:}

This corresponding to the measurements of 
the expansion rate of the universe, $H(z)$, in the redshift range 
$0.07 \leq z \leq 2.34$ \cite{Morescoetal2012,Morescoetal2016,Moresco2015}. These data points are retrieved using two strategies:
\begin{itemize}
    \item[1] \underline{Differential age technique}: In this technique, the data-points are calculated using the relation between the redshift $z$ and the rate of change of the galaxy's age.  
    \begin{equation}
         \label{dtdz}
        \frac{dt}{dz} =\, -\, \frac{1}{(1 + z) H(z)} \,\,.
    \end{equation}
    \item[2] \underline{Galaxy clustering technique}: The data points are obtained by utilising galaxy or quasar clustering that provides direct measurements of $H(z)$ from the radial peaks of baryon acoustic oscillations (BAO) 
    \cite{Blake:2011en,ballardini3}. 
    
The $\chi^2$ for the Hubble parameter measurements is
\begin{equation}
\chi^2_H=\sum_{i}\bigg[\frac{H_i^{th}-H_i^{obs}}{\sigma^{H}_i}\bigg]^2 \,.
\end{equation}
\end{itemize}

\paragraph{\textbf{Masers Data}:}

The Megamaser Cosmology Project's measures the angular diameter distances.
Its $\chi^2_{\rm{masers}}$ is defined as
\begin{equation}
\chi^2_{\rm{mas}}=\sum_{i}\bigg[\frac{D_{Ai}^{th}-D_{Ai}^{obs}(z_i)}{\sigma^{D}_i}\bigg]^2
\end{equation}
such that
\begin{equation}
D_A(z)=\frac{1}{1+z}\int_0^z\frac{dz^\prime}{H(z^\prime)}
\end{equation}
where $D_A^{th}$ is the angular diameter distance.

\section{Parametric estimations}

The Bayesian inference approach, which is frequently employed for parameter estimation in cosmological models, is used here to carry out the statistical 
analysis. This statistics states that the posterior probability
distribution function of model parameters is directly proportional to the prior probability of the model parameters and their likelihood function. 
The likelihoods used to estimate the parameters are multivariate Gaussian likelihoods and are given by
\begin{equation}
\mathcal{L} (\Theta)\propto {\rm exp}\left[-\frac{\chi^2(\Theta)}{2}\right] \,,
\end{equation}
where $\Theta$ belongs to a set of parameters. The posterior probability is proportional to ${\rm exp}[-\frac{\chi^2(\Theta)}{2}]$. Consequently, a minimum $\chi^2(\Theta)$ will guarantee a maximum likelihood. In our analysis the Gaussian likelihood is given by
\begin{equation}
\mathcal{L} \propto {\rm exp}\left[-\frac{\chi^2_{T}}{2}\right] \,,
\end{equation}
where $\chi^2_{T}: = \chi^2_{\rm SN}+\chi^2_{\rm H}+\chi^2_{\rm masers}$.
This formulation will be same for both the parametrizations. For the Polynomial parametrization the parameters that need to be constrained are: $\alpha$, $\beta$ and $\widetilde{h}$, whereas, for the exponential parametrization we have $\widetilde{h}$ and $\alpha$. In our analysis, we employed uniform priors. For the estimations, the technique that is used is 
Markov Chain Monte Carlo (MCMC). The obtained MCMC chains then is studied 
using the GetDist program \cite{Lewis}. 

\section{Constraints on Hubble parameter and dark energy equation of state 
parameter: Phantom crossing and Hubble tension}

\begin{figure*}[!ht]
\centering
\includegraphics[scale=0.8]{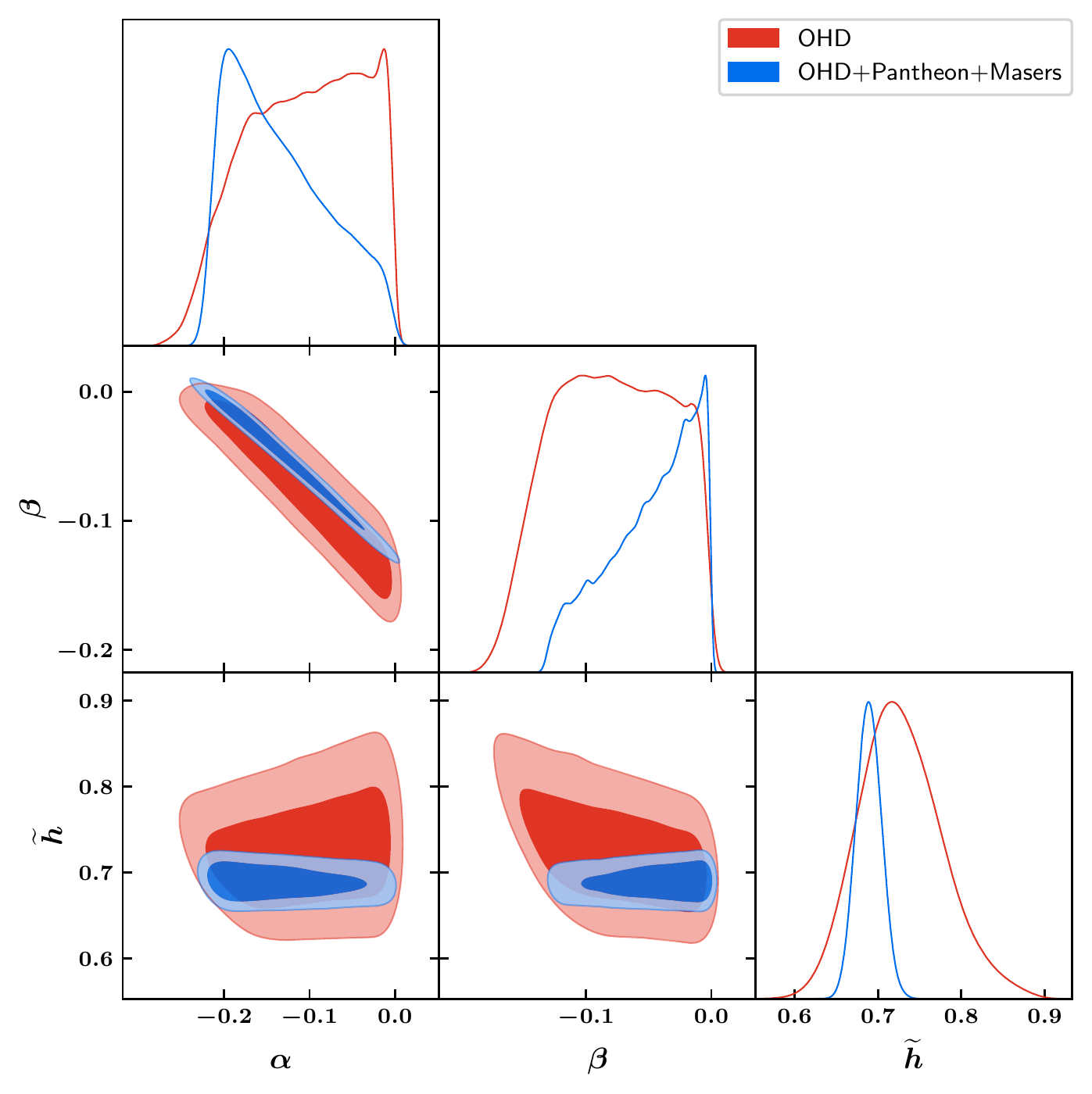}
\caption{Polynomial: $2\sigma$ contour levels between $\alpha$, $\beta$ and $\widetilde{h}$ for OHD and its combinations with Pantheon+Masers 
\cite{shah}.}
\label{fig: poly_contour}
\end{figure*}

\begin{figure*}[!ht] 
\centering
\begin{subfigure}{0.495\linewidth} \centering 
   \includegraphics[height=5.4cm,width=6.4cm]{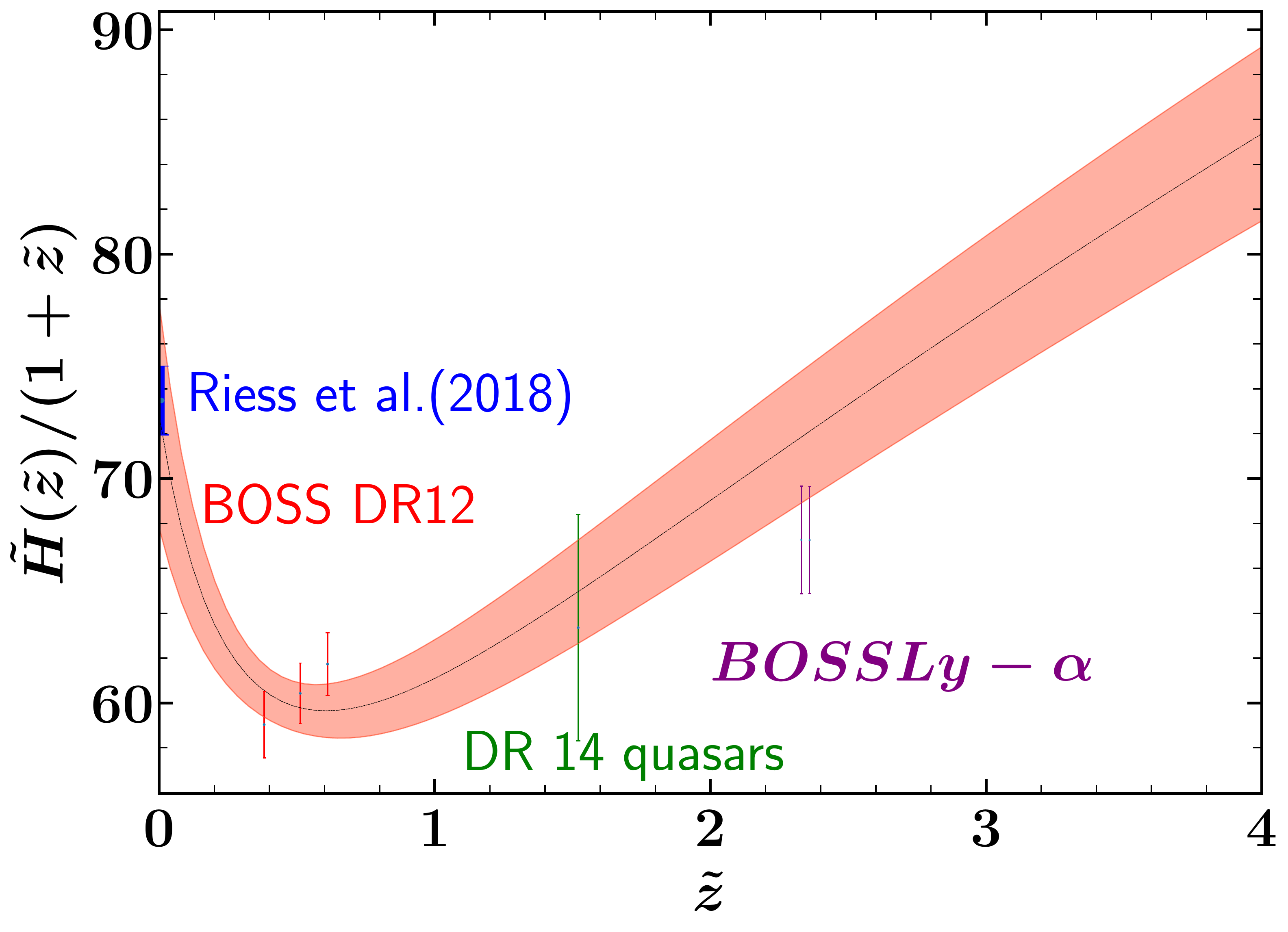}
   \caption{} \label{poly-H-a}
\end{subfigure}
\begin{subfigure}{0.495\linewidth} \centering
    \includegraphics[height=5.4cm,width=6.4cm]{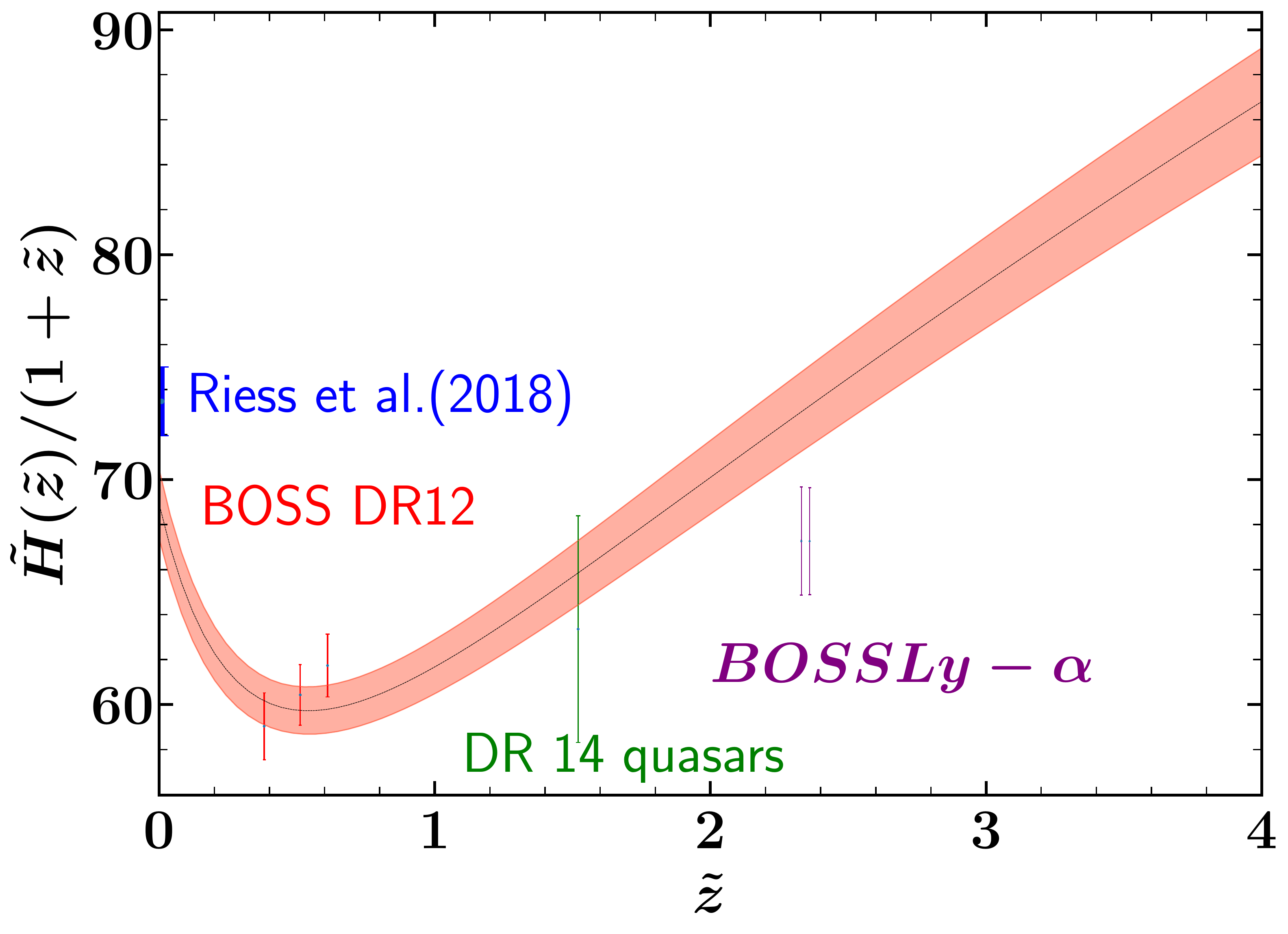}
    \caption{} \label{poly-H-b}
\end{subfigure}
\caption{Polynomial: Figures (\ref{poly-H-a}) and (\ref{poly-H-b}) depicts the evolution of $\widetilde{H}(\widetilde{z})/(1+\widetilde{z})$ with $\widetilde{z} \in [0,4]$ for the datasets OHD and OHD+Pantheon+Masers, respectively. The dark line represents the best-fit and the shaded region corresponds to the $1\sigma$ limit 
\cite{shah}. }
\label{fig:H_poly}
\end{figure*}
%%%%%%%%%%%%%%%%%%%%%%%
\begin{figure*}[!ht] 
\centering
\begin{subfigure}{0.495\linewidth} \centering 
   \includegraphics[height=5.4cm,width=7cm]{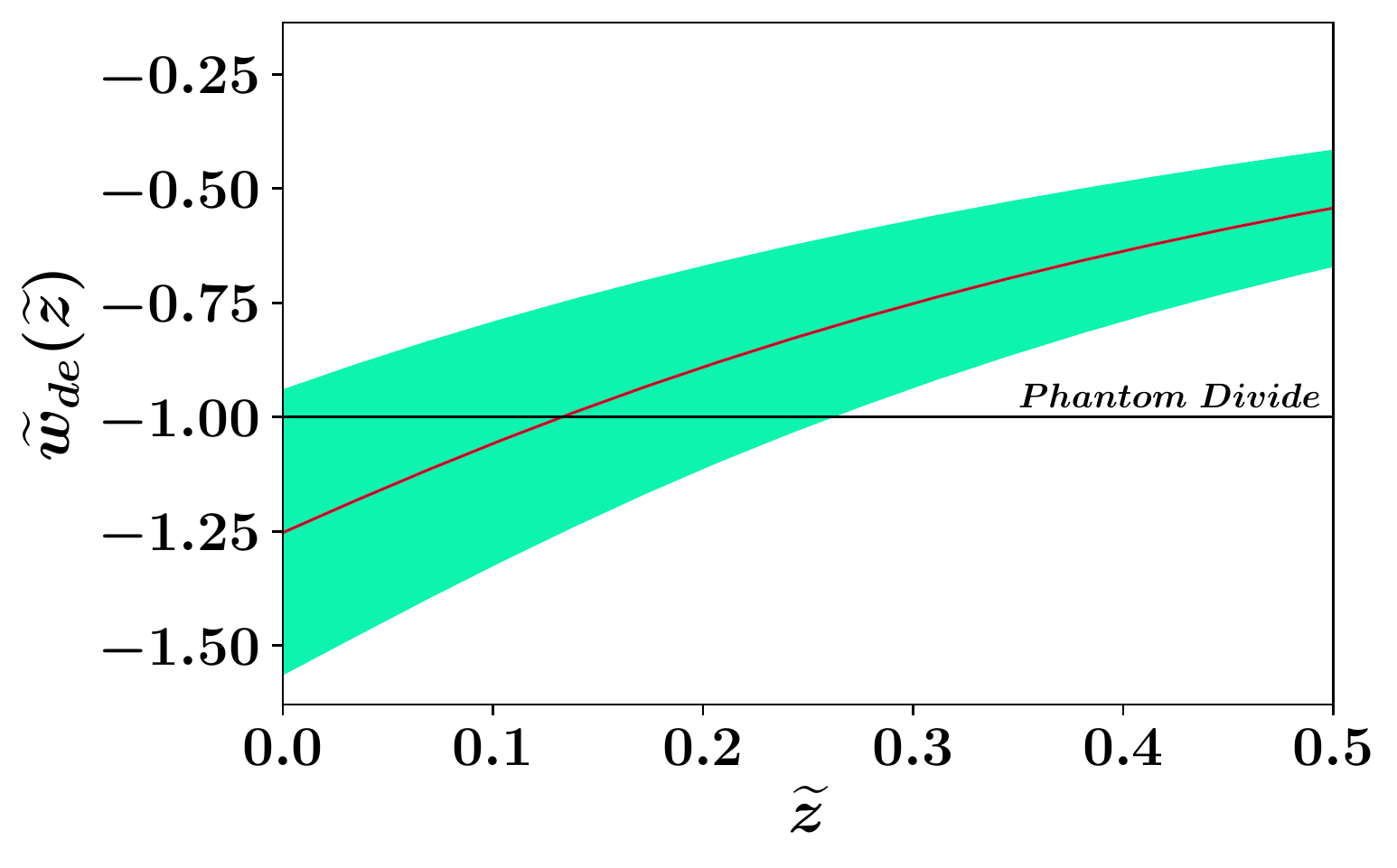}
   \caption{} \label{poly-wde-a}
\end{subfigure}
\begin{subfigure}{0.495\linewidth} \centering
    \includegraphics[height=5.4cm,width=7cm]{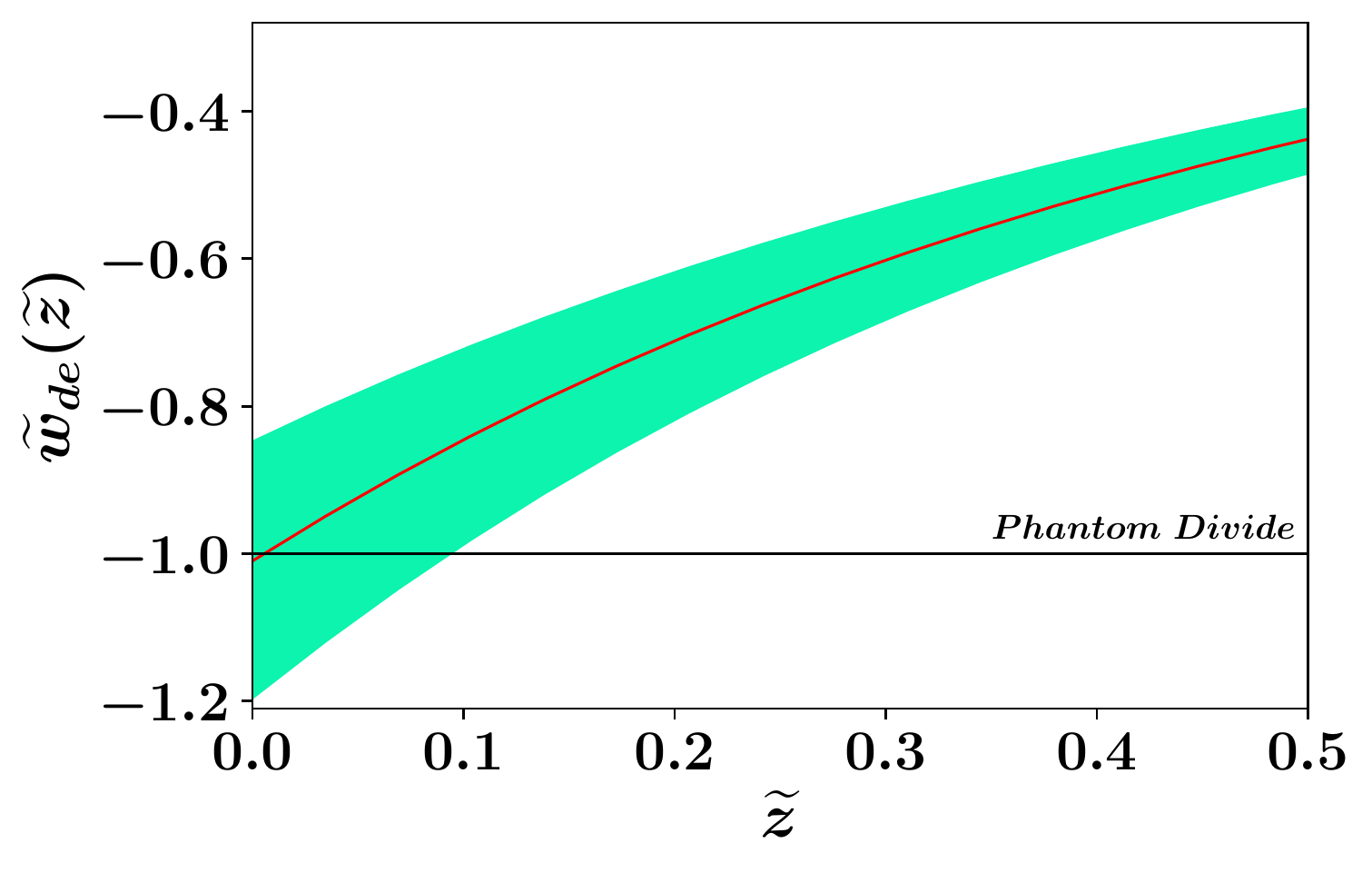}
    \caption{} \label{poly-wde-b}
\end{subfigure}
\caption{Polynomial: Figures (\ref{poly-wde-a}) and (\ref{poly-wde-b}) depicts the evolution of $\widetilde{w}_{de}(\widetilde{z})$ with $\widetilde{z} \in [0,4]$ for the datasets OHD and OHD+Pantheon+Masers, respectively. The dark line represents the best-fit and the shaded region corresponds to the $1\sigma$ limit \cite{shah}
%All the datasets allows the Phantom crossing within $1\sigma$ limit.
}
\label{fig:wde_poly}

\end{figure*}

From the obtained chains of parameters $\alpha$, $\beta$ and $\widetilde{h}$, 
from our MCMC simulation, upto $2\sigma$ (shown in fig.\,\ref{fig: 
poly_contour}). In fig.\,(\ref{fig: poly_contour}), we have plotted the 
obtained parametric dependence between $\alpha$, $\beta$ and $\widetilde{h}$ 
using OHD and its combination with Pantheon and Masers. From this figure, one 
can note that the combination of all data set significantly reduces the errors
bars on $\widetilde{h}$ as compared to only OHD data set. The obtained results
are given in table (\ref{table1}) where we show that only OHD data set give  
$\widetilde{h}$ significantly larger than the combined data set. In 
particular, we have not found any significant Hubble tension for OHD, even for
the combined data set, the tension reduces to $1.3\sigma$ level. The reduction of this tension can be attributed to the phantom crossing taking place in case of the polynomial parametrization. 

\begin{table*}[ht]
\centering
\renewcommand{\arraystretch}{1.2}
\begin{tabular}{|c|c||c|c|}  %
	\hline
	&   \multicolumn{2}{|c|}{Parametrizations} & \\
	\cline{2-3}
	Observational &  Polynomial & Exponential & $\Lambda$CDM \\
	dataset &  Best-fit($\pm 1\sigma$) &  Best-fit($\pm 1\sigma$) & \\
	\hline
	  & $\widetilde{h} =  0.7279_{-0.05}^{+0.05}$  & $\widetilde{h} = 0.671_{-0.029}^{+0.029} $  & $\widetilde{h} = 0.6770^{+0.030}_{-0.030}$  \\
 	 OHD & $ \alpha=-0.101^{+0.07}_{-0.077}$ &  $\alpha=-0.299^{+0.043}_{-0.042}$ & $\widetilde{\Omega}_M = 0.3249^{+0.064}_{-0.059}$  \\
	 & $\beta = -0.078^{+0.051}_{-0.049}$ &  - & \\
	\hline
	 & $\widetilde{h}$ =  $0.689_{-0.015}^{+0.015}$  & $\widetilde{h}$ = $0.677_{-0.007}^{+0.007} $ & $\widetilde{h}=0.6683^{+0.026}_{-0.026}$\\
OHD+Pantheon +Masers	& $\alpha= -0.145^{+0.078}_{-0.051}$   & ${\alpha}$ =  $-0.335_{-0.017}^{+0.016}$ & $\widetilde{\Omega}_M = 0.3440^{+0.061}_{-0.054}$\\
& $\beta = -0.041^{+0.029}_{-0.047}$ &  - & \\
	\hline\hline
\end{tabular}
\caption{Best-fits with their $1\sigma$ levels for polynomial and exponential parametrizations, and for the $\Lambda$CDM model from OHD and OHD+Pantheon+Masers datasets \cite{shah}. }
\label{table1}
\end{table*}

The corresponding evolution of $\widetilde{w}_{de}(\widetilde{z})$ and $\widetilde{H}(\widetilde{z})$ upto $1\sigma$ are shown in figures (\ref{fig:H_poly}) and (\ref{fig:wde_poly}), respectively. In the fig (\ref{fig:H_poly}), it can be seen that because of the large $1\sigma$ deviations, the OHD data set alone does not show a tension with the 
Riess et. al. \cite{riess2018} and BOSSLy-$\alpha$ \cite{Ly1a}.
While the combined data set exhibits a substantial tension with the 
Riess et al. by yielding a comparably smaller $\widetilde{h}$ (\ref{fig:H_poly}).
Additionally, we demonstrate in fig.\,(\ref{fig:wde_poly}) that both 
data sets result in a phantom crossing near the current epoch. 
It is noteworthy that this property arises exclusively as a result of the coupling between two matter components (BM+DM), and with no additional 
degrees of freedom. 

%%%%%%%%%%%%%%%%%%% Exponential %%%%%%%%%%%%%%%%%%%%%%%%

\begin{figure*}[!ht]
\centering
\includegraphics[scale=0.86]{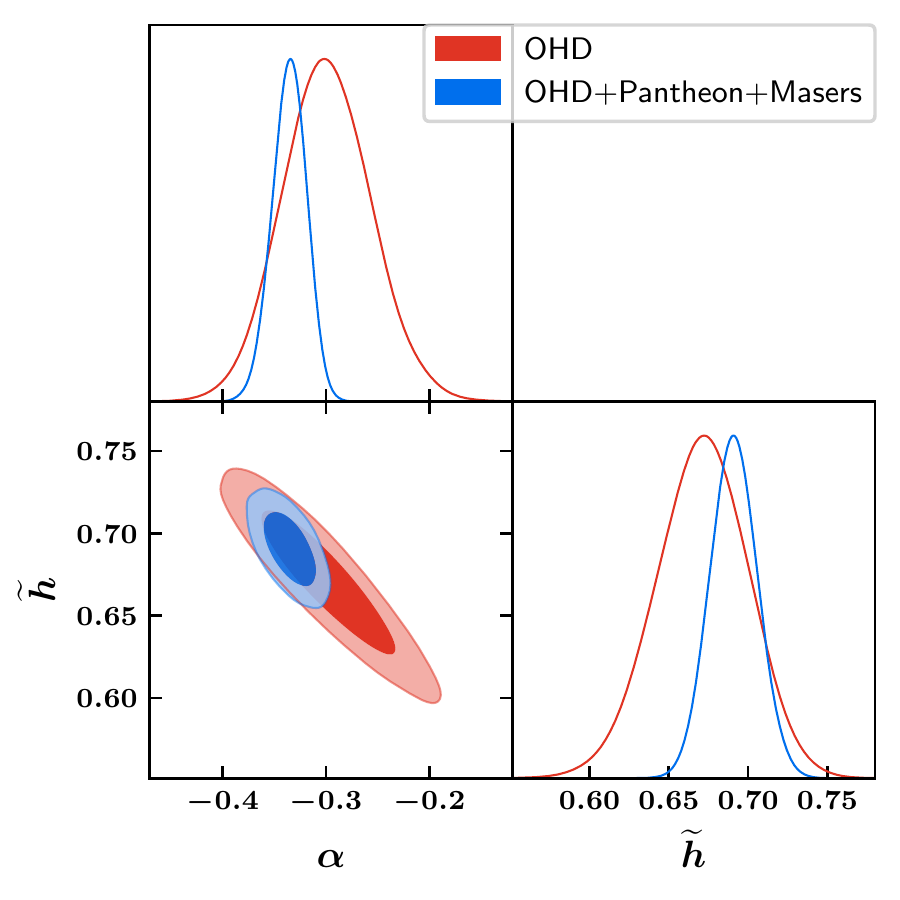}
\caption{Exponential: $2\sigma$ contour levels between $\alpha$ and $\widetilde{h}$ for OHD and its combinations with Pantheon+Masers \cite{shah}.}
\label{fig:tri_exp}
\end{figure*}

Similar analysis is also carried out for the second parametrization, and in this instance, we demonstrate the parametric dependency between $\alpha$ and $\widetilde{h}$ for two sets of data in fig. (\ref{fig:tri_exp}) and the resulted constraints are shown in the table (\ref{table1}). Here, the value of the Hubble constant is consistent rather with the $DES+ BAO+ Planck$ combined data upto the $1$-$\sigma$ level (see table (2)) and Eq.\,(45) of \cite{Planck18}). As a result, the tension does not significantly decrease. This is to be expected because the exponential parametrization model mimics the 
$\Lambda$CDM since there is no phantom crossover (prior to the current epoch). 

\begin{figure*}[!ht] 
\centering
\begin{subfigure}{0.495\linewidth} \centering 
   \includegraphics[height=5.6cm,width=6.8cm]{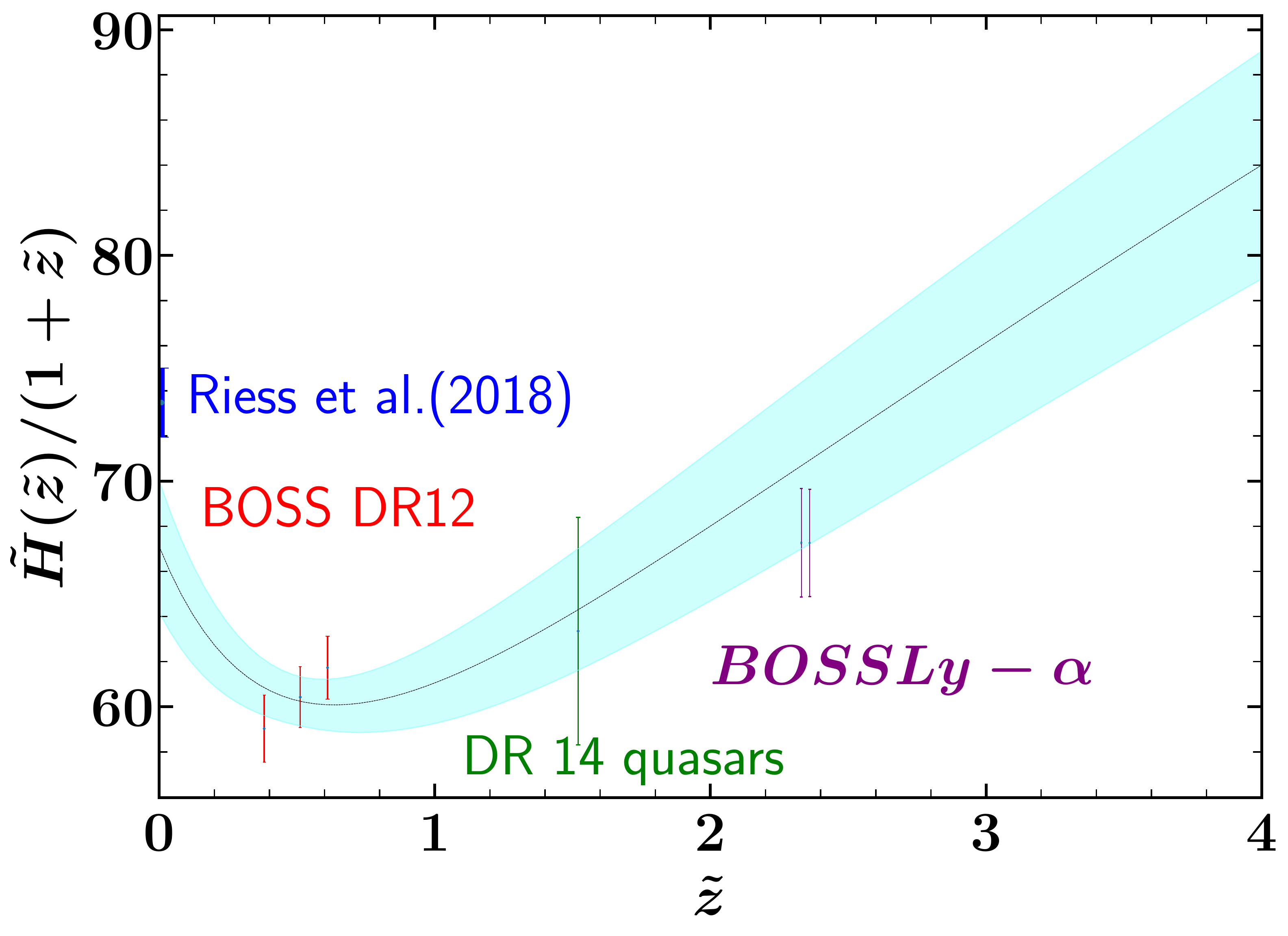}
   \caption{} \label{exp-H-a}
\end{subfigure}
\begin{subfigure}{0.495\linewidth} \centering
    \includegraphics[height=5.6cm,width=6.8cm]{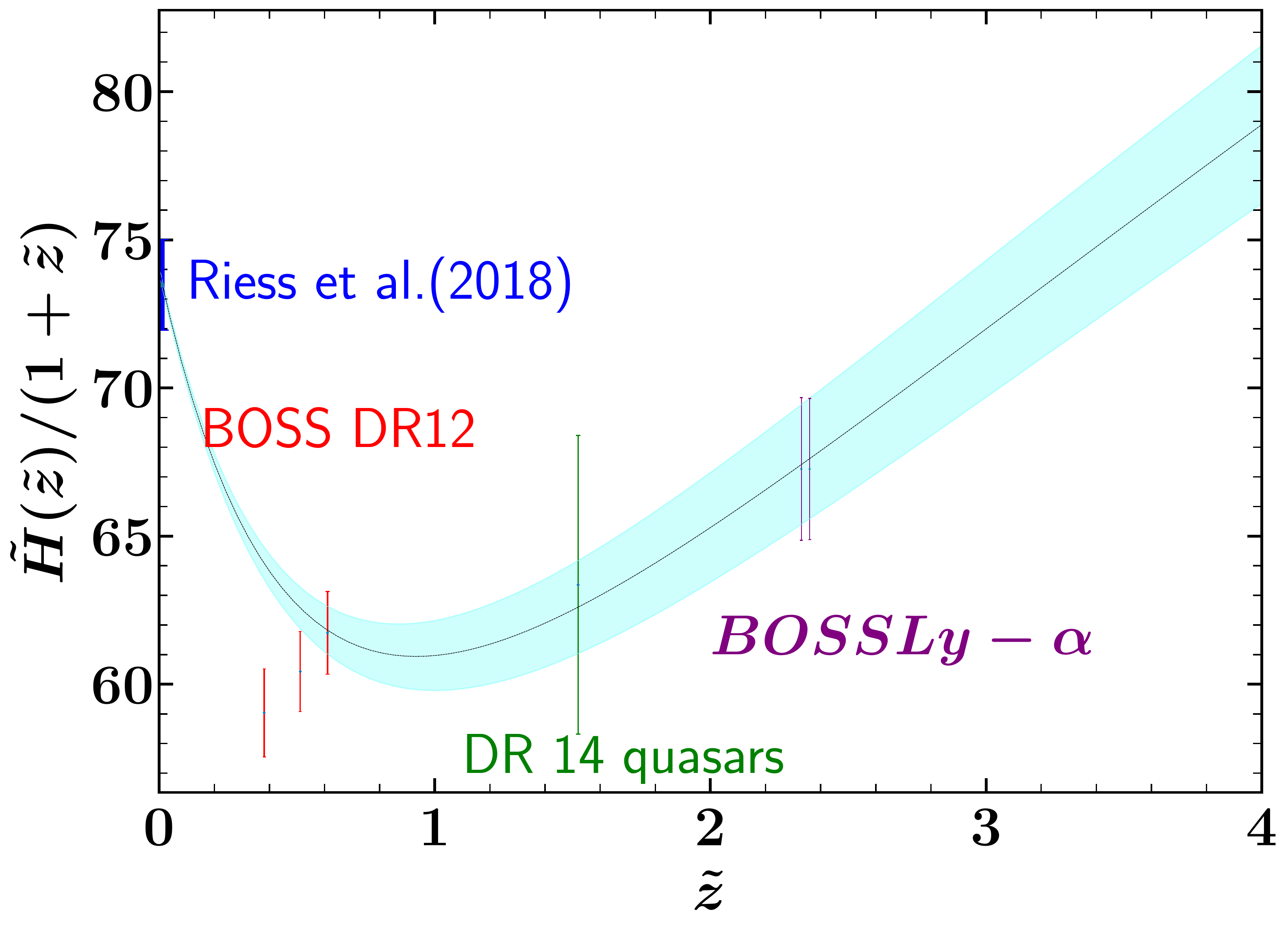}
    \caption{} \label{exp-H-b}
\end{subfigure}
\caption{Exponential: Figures (\ref{exp-H-a}) and (\ref{exp-H-b}) depicts the evolution of $\widetilde{w}_{de}(\widetilde{z})$ with $\widetilde{z} \in [0,4]$ for the datasets OHD and OHD+Pantheon+Masers, respectively. The dark line represents the best-fit and the shaded region corresponds to the $1\sigma$ limit \cite{shah}.}
\label{fig:figure3}
\end{figure*} 
%%%%%%%%%%%%%%%%%%%%%%%
\begin{figure*}[!ht] 
\centering
\begin{subfigure}{0.495\linewidth} \centering 
   \includegraphics[height=5.4cm,width=8cm]{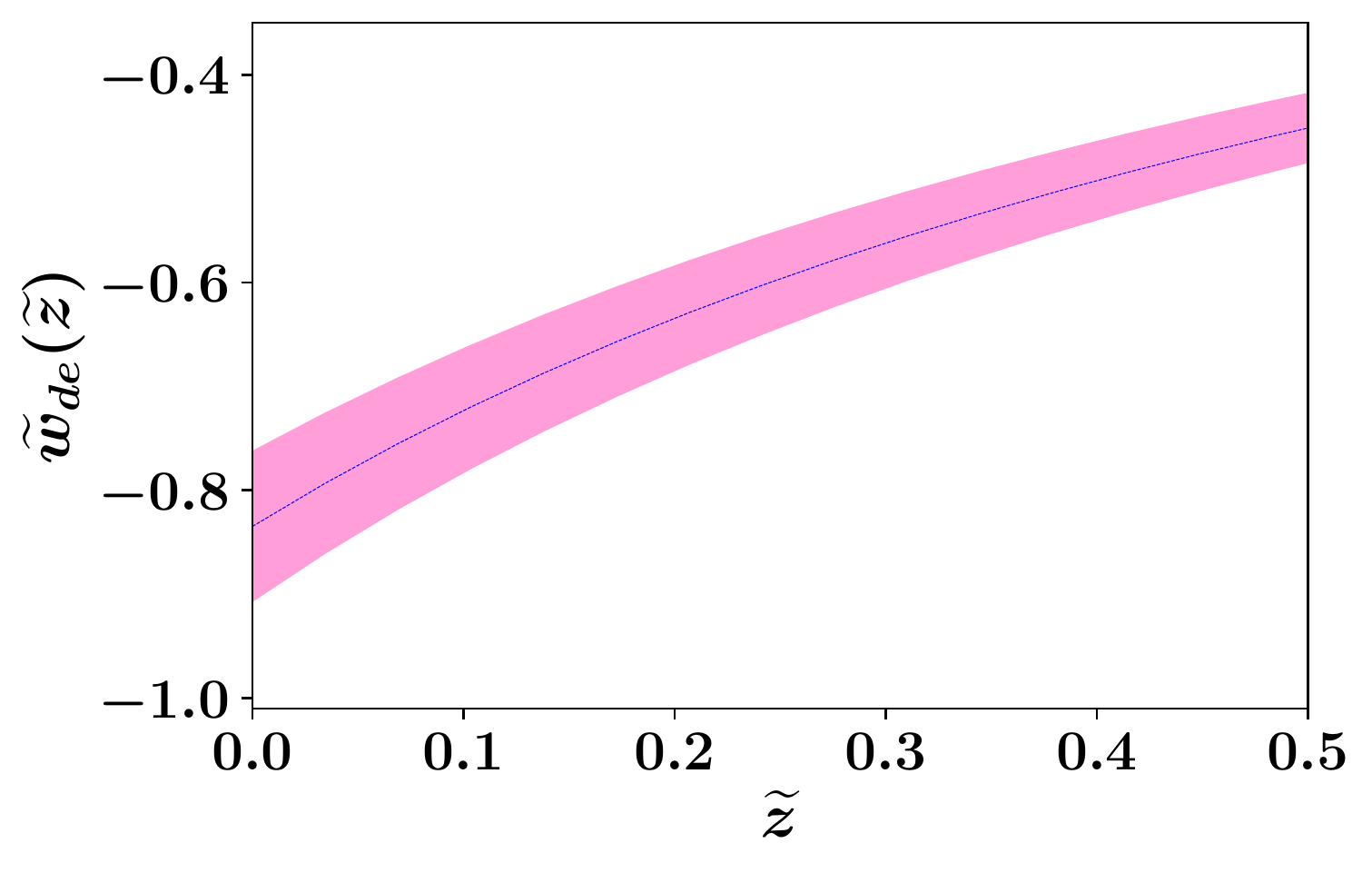}
   \caption{} \label{exp-wde-a}
\end{subfigure}
\begin{subfigure}{0.495\linewidth} \centering
    \includegraphics[height=5.4cm,width=8cm]{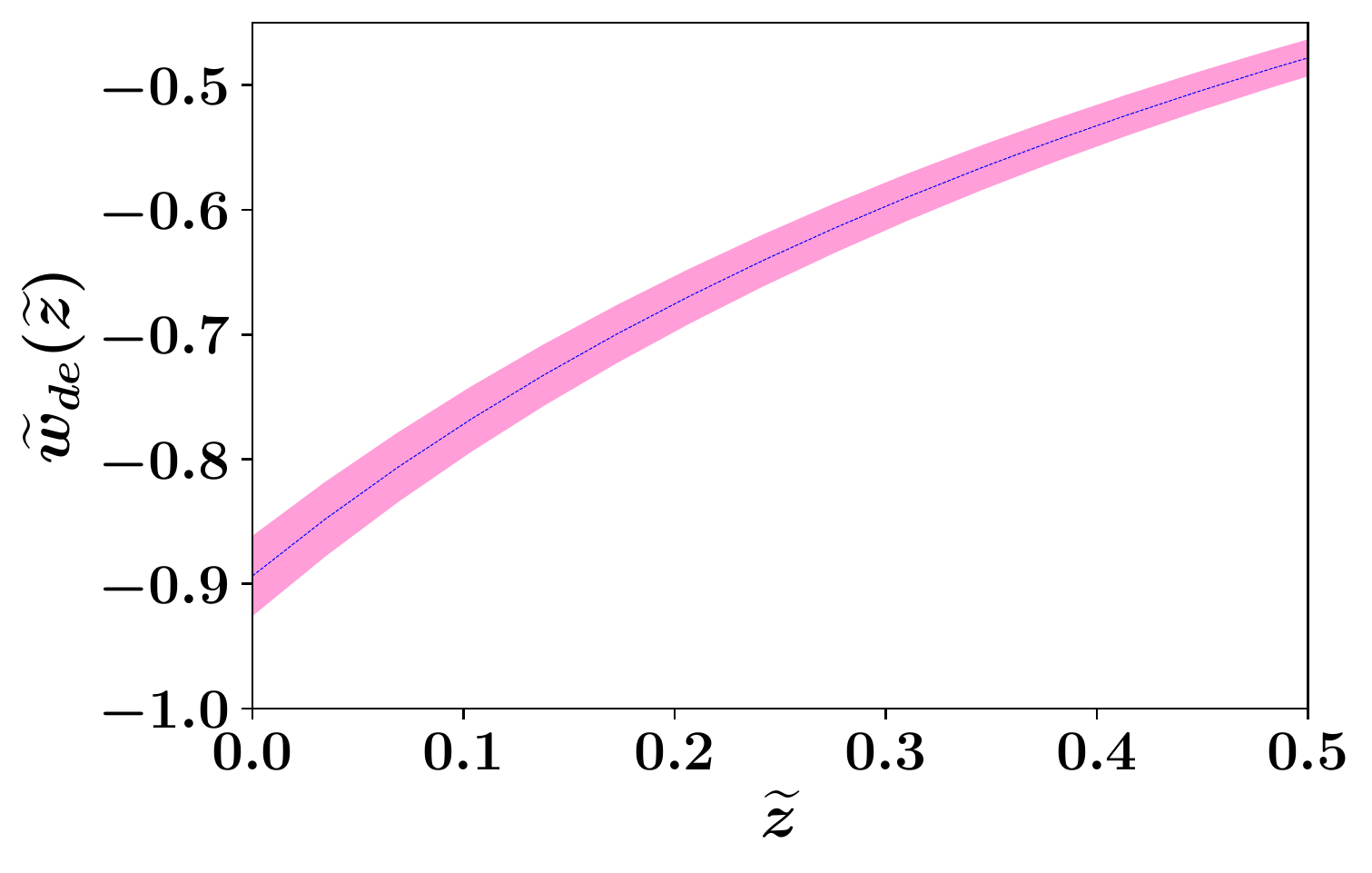}
    \caption{} \label{exp-wde-b}
\end{subfigure}
\caption{Exponential: Figures (\ref{exp-wde-a}) and (\ref{exp-wde-b}) depicts the evolution of $\widetilde{w}_{de}(\widetilde{z})$ with $\widetilde{z} \in [0,4]$ for the datasets OHD and OHD+Pantheon+Masers, respectively. The dark line represents the best-fit and the shaded region corresponds to the $1\sigma$ limit \cite{shah}.
}
\label{fig:wde_exp}
\end{figure*}
%%%%%%%%%%%%%%%%%%%%%%%%%
%%%%%%%%%%%%%%%%%%%%%%
\begin{figure*}[!ht] 
\centering
\begin{subfigure}{0.495\linewidth} \centering 
   \includegraphics[height=5.5cm,width=8cm]{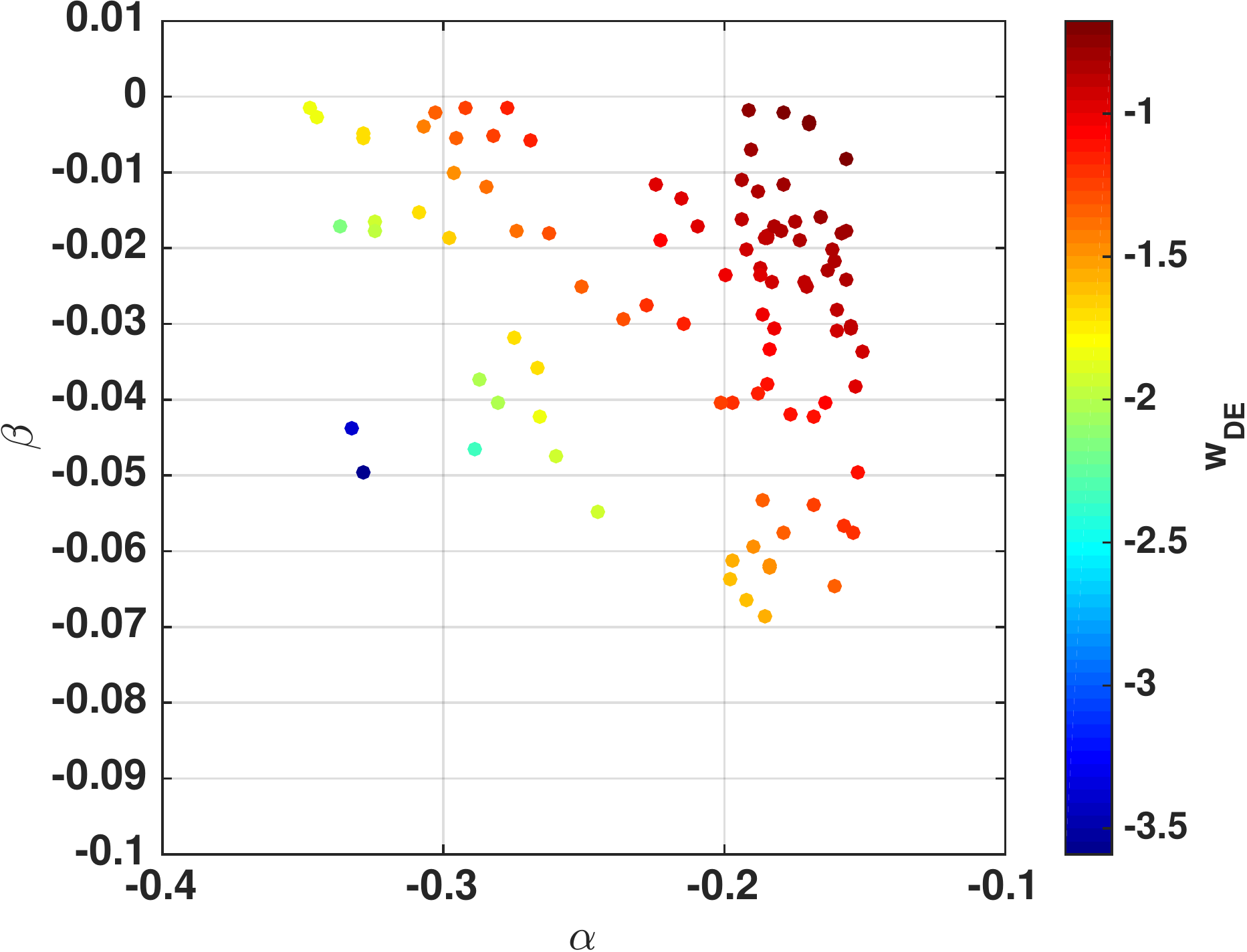}
   \caption{} \label{polyparam}
\end{subfigure}
\begin{subfigure}{0.495\linewidth} \centering
    \includegraphics[height=5.5cm,width=8cm]{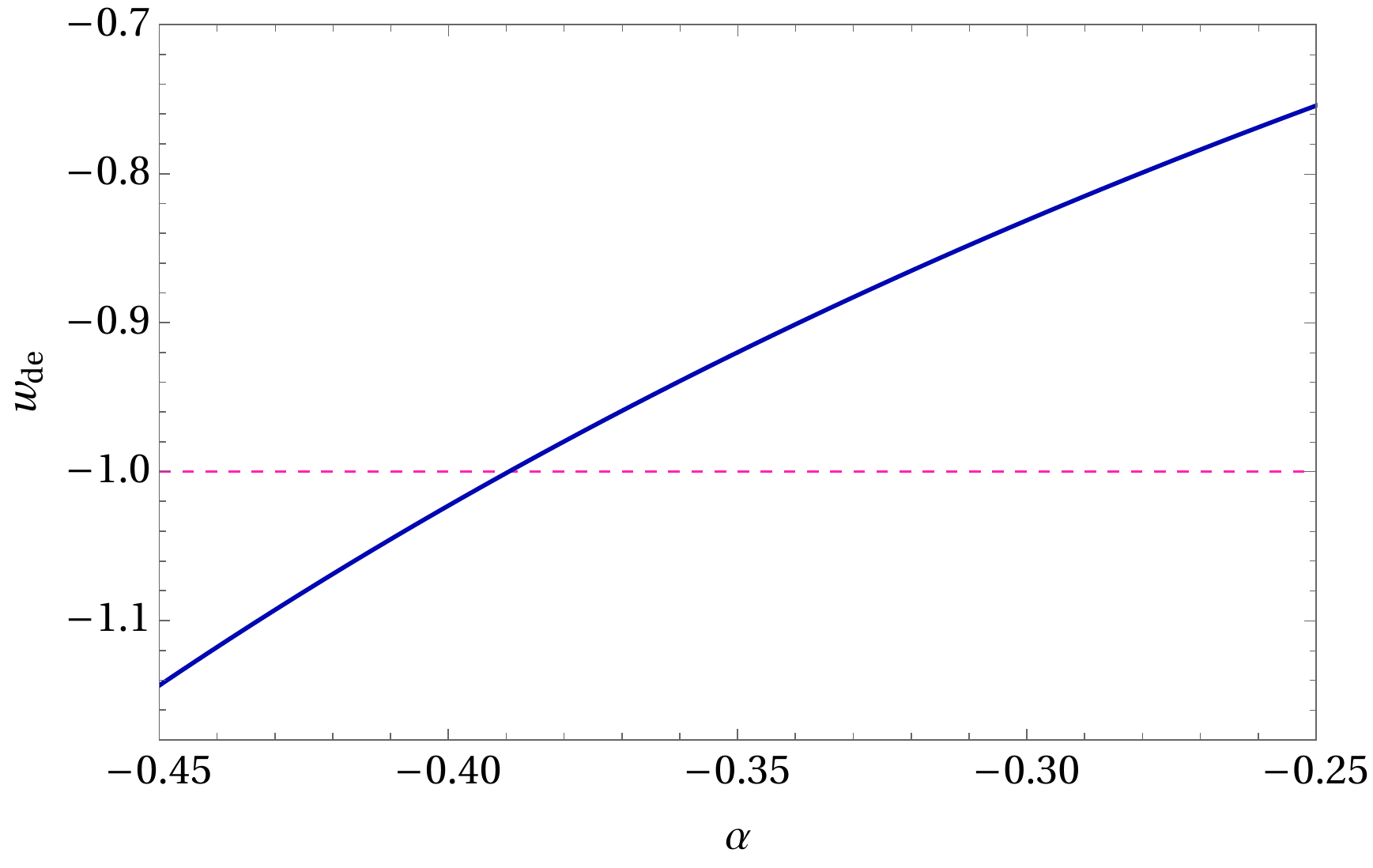}
    \caption{} \label{expparam}
\end{subfigure}
\caption{Exponential: Figures (\ref{polyparam}) and (\ref{expparam}) shows the variation of the two model parameters with equation of state of the dark energy ($w_{DE}$) \cite{shah}}
\label{fig:param_wde}

\end{figure*}

In contrast to the polynomial scenario, the combined data set agrees with 
Riess et al. and BOSSLy-$\alpha$, which is not the case when utilising 
simply OHD data set, as can be shown in fig.\,(\ref{fig:figure3}). 
Additionally, it can be seen from fig.\,(\ref{fig:wde_exp}) that the phantom 
crossing of $\widetilde{w}_{de}(\widetilde{z})$ does not occur. 

Moreover, from the fig.\,(\ref{fig:wde_exp}), one can notice that the phantom 
crossing of $\widetilde{w}_{de}(\widetilde{z})$ does not happen. In 
fig.\,(\ref{fig:param_wde}), we explicitly show the dependence of model parameters on the $\widetilde{w}_{de}$. 
In fig.\,(\ref{polyparam}), we demonstrate that both $\alpha$ and $\beta$ must 
be negative in order to produce the phantom crossing. Similarly, for the 
exponential case, fig.\,(\ref{fig:wde_exp}) illustrates the dependency of 
$\widetilde{w}_{de}$ on parameter $\alpha$.

\section{Comparison with $\Lambda$CDM}

Let us first note that while comparing the two parametrizations, polynomial (\ref{m1}) and exponential (\ref{m2}), with the $\Lambda$CDM model, the 
inclusion of the additional parameters in comparison to the standard model must be taken into consideration. 

To compare the two parametrizations, polynomial (\ref{m1})) and exponential( (\ref{m2})) with the vanilla $\Lambda$CDM model, one needs to take care of the introduction of the extra parameters with respect to the standard model. We have two extra parameters for polynomial and one extra parameter for exponential.
In order to handle it, a thorough Bayesian Information Criterion (BIC) is calculated. In BIC analysis, a model with more parameters gets 
penalised more. Under the assumption that the model errors are independent and obey a normal distribution, then the BIC can be rewritten in terms of $\Delta \chi^2$ as 
\begin{equation}
BIC\approx \Delta \chi^2 + df. \ln(n)
\end{equation}
where $df$ is the number of free parameters in the test and $n$ is the number of points in the observed data. In the Table\,(\ref{tablebic}), we provide details about our findings. The polynomial parametrization has good evidence, as can be shown from Table\,(\ref{tablebic}), when compared to the typical $\Lambda$CDM case.  Any indication that $\Delta BIC \geq 10$ indicates extremely strong support for the novel model proposed in comparison to the conventional one. 
Even while the exponential scenario has strong support when the data are combined, it has no such support when the OHD data are the only ones taken into account. We find compelling support for the polynomial parametrization for both OHD and combined data. 
\begin{table}[htb]
\centering
\begin{tabular}{|l|l|l|l|l|}
\hline
\begin{tabular}[c]{@{}l@{}}Observational \\      Dataset\end{tabular} & \begin{tabular}[c]{@{}l@{}}Polynomial\\   ($\Delta$ BIC)\end{tabular} & \begin{tabular}[c]{@{}l@{}}Polynomial\\ Evidence\end{tabular} & \begin{tabular}[c]{@{}l@{}}Exponential\\   ($\Delta$ BIC)\end{tabular} & \begin{tabular}[c]{@{}l@{}}Exponential\\   Evidence\end{tabular} \\ \hline
OHD                                                                   & 9.63                                                                   & Strong                                                        & 0.62                                                                  & Not worth                                                       \\ \hline
OHD+Pantheon +Masers                                                  & 8.88                                                                   & Strong                                                        & 4.01                                                                  & Positive                                                        \\ \hline
\end{tabular}
\caption{The evidence in support of polynomial and exponential parametrizations for OHD and OHD+Pantheon+Masers datasets with respect to the standard $\Lambda$CDM scenario \cite{shah}. }
\label{tablebic}
\end{table}

\section{Conclusion and future perspectives }
\label{conclusion}

In this paper, we have shown the observable validity and wider implications of the dark matter and baryonic matter interaction in the Einstein frame, which is produced by a general disformal transformation between the Jordan and the Einstein frames. The idea behind the phenomena is that dark matter adheres to Einstein frame geodesics whereas baryonic matter follows Jordan frame trajectories. Since both matter components are coupled together under the usual disformal transformation, their respective energy conservation in the Einstein frame is lost. 

We employ two distinct parametrizations to relate the scale factors of both frames in the conventional FRW space-time since the geodesics of the two frames are not identical (as a result of the disformal transformation between them). To obtain the constraints on the model parameters in the Jordan frame, which we assume to be the physical frame, we specifically use the polynomial and exponential parametrizations (the Jordan frame is used for all observations because the underline mechanism predicts that baryonic matter will follow its paths in this frame).
The best-fit Hubble parameter values for two distinct data combinations are such that they significantly lessen the so-called ``Hubble tension'' in the case of the polynomial parametrization Eq.\,(\ref{m1}). In the case of polynomial parametrization, the tension for OHD data is negligible, with a computed value of $\tilde{h}=0.7279^{+0.05}_{-0.05}$ $1-\sigma$ consistent with Riess et al. However, with the combined OHD + Pantheon + Masers data, the tension is lowered to $1.3-\sigma$ level.  
We would like to emphasize that this is associated to the fact that, 
in this specific parametrization, the dark energy equation of state passes from quintessence to phantom regime. Thus, when doing the $\Delta$BIC analysis, which is shown in Table\,(\ref{tablebic}), 
we notice substantial evidence in favour of this parametrization. 

As for the exponential parametrization Eq.\,(\ref{m2}), we have not noticed any appreciable Hubble-tension reduction (see, Fig.\,(\ref{fig:tri_exp})) with both the data combinations and it is understood by the model's quintessence-like behaviour with $\widetilde{w} _{de}(\widetilde{z})\geq -1$ around the current epoch. In the exponential case, we only find a little amount of positive evidence in the combined data scenario, which is not substantial enough when simply taking OHD data into account (see Table\,(\ref{tablebic})).

Let us again emphasise that the scenario under consideration which is based on the interaction between DM and BM enables late-time cosmic acceleration without including any exotic fluid and is also consistent with observations. The occurrence of phantom crossover, which is currently supported by the majority of observations, is one of the most significant and general implications of the DM-BM interaction.
Last but not least, given the existence of disformal coupling between DM and BM, it would be intriguing to take perturbations into account and examine the matter power spectrum which we will try to report soon.


\begin{thebibliography}{99}

%%%%%%%%%%%%%%%%%%%%%%%%% Reviews: Dark Energy %%%%%%%%%%%%%%%%%%%%%%

\bibitem{DEbook}
L. Amendola and S. Tsujikawa, {\it Dark energy. Theory and observations}, Cambridge Univ. Press, Cambridge (2010). 

\bibitem{Copeland} 
E.~J.~Copeland, M.~Sami and S.~Tsujikawa, ``Dynamics of dark energy'',
Int.\ J.\ Mod.\ Phys.\ D \textbf{15}, 1753 (2006) 

\bibitem{Sami} 
M.~Sami, ``Why is Universe so dark?'',New Adv.\ Phys.\ \textbf{10}, 77 (2016), arXiv:1401.7310 [physics.pop-ph]. 

\bibitem{Sami2} 
M.~Sami, ``A Primer on problems and prospects of dark energy,''
Curr.\ Sci.\ \textbf{97}, 887 (2009), arXiv:0904.3445 [hep-th]. 

\bibitem{Sami-notes} 
M.~Sami, ``Models of dark energy,``Lect.\ Notes Phys.\ \textbf{720}, 219 (2007); Nur Jaman, M. Sami,  Galaxies 10 (2022) 2, 51, e-Print:2202.06194 [gr-qc]; M. Sami, R. Myrzakulov, Gen.Rel.Grav. 54 (2022) 8, 86; M. Sami, 
R. Gannouji,  Int.J.Mod.Phys.D 30 (2021) 13, 2130005, arXiv: 2106.00843 [gr-qc];  M. Sami, Pravabati Chingangbam    Phys.Rev.D 66 (2002) 043530.
\bibitem{Trodden} 
A.~Silvestri and M.~Trodden, ``Approaches to Understanding Cosmic Acceleration,''Rept.\ Prog.\ Phys.\ \textbf{72}, 096901 (2009), arXiv:0904.0024 [astro-ph.CO]. 

\bibitem{Sahni} 
V.~Sahni and A.~A.~Starobinsky, ``The Case for a positive cosmological Lambda term'', Int.\ J.\ Mod.\ Phys.\ D \textbf{9}, 373 (2000), astro-ph/9904398. 

\bibitem{Frieman} 
J.~Frieman, M.~Turner and D.~Huterer, ``Dark Energy and the Accelerating Universe'', Ann.\ Rev.\ Astron.\ Astrophys.\ \textbf{46}, 385 (2008) 
arXiv:0803.0982 [astro-ph]. 

\bibitem{Caldwell} 
R.~R.~Caldwell and M.~Kamionkowski, ``The Physics of Cosmic Acceleration'',
Ann.\ Rev.\ Nucl.\ Part.\ Sci.\ \textbf{59}, 397 (2009) 
arXiv:0903.0866 [astro-ph.CO]. 

\bibitem{Peri} 
L.~Perivolaropoulos, ``Accelerating universe: observational status and theoretical implications,''AIP Conf.\ Proc.\ \textbf{848}, 698 (2006), astro-ph/0601014. 

\bibitem{Frieman2} 
J.~A.~Frieman, ``Lectures on dark energy and cosmic acceleration,''
AIP Conf.\ Proc.\ \textbf{1057}, 87 (2008), arXiv:0904.1832 [astro-ph.CO].

\bibitem{Carroll} 
S.~M.~Carroll, ``The Cosmological constant,''Living Rev.\ Rel.\ \textbf{4}, 1 (2001), astro-ph/0004075. 

\bibitem{Pady} 
T.~Padmanabhan, ``Cosmological constant: The Weight of the vacuum,''
Phys.\ Rept.\ \textbf{380}, 235 (2003), hep-th/0212290. 

%%%%%%%%%%%%%%%%%%%%%%%%%%%% Age %%%%%%%%%%%%%%%%%%%%%%%%%%%%%%%%%%%%%%

\bibitem{age}
Spergel, D. N., Bolte, M. and Freedman, W. ,{\it The age of the universe},Proceedings of the National Academy of Sciences, {\bf 94}, 6579-6584(1997).

\bibitem{age2}
J. Dunlop, J. Peacock, H. Spinrad, A. Dey, R. Jimenez, D. Stern and R. Windhorst ,{\it A 3.5-Gyr-old galaxy at redshift 1.55},Nature {\bf 381},581–584(1996).
\bibitem{Damjanov}
I. Damjanov, et al. {\it Red Nuggets at $z\sim 1.5$: Compact passive galaxies and the formation of the Kormendy Relation}, The Astrophysical Journal  {\bf 695} 1 (2009): 101.

\bibitem{age3}
M. Salaris, S. Degl'Innocenti, And A. Weiss, {\it The Age of the Oldest Globular Clusters},The Astrophysical Journal, {\bf 479},665–672(1997).
\bibitem{LopezCorredoira2017}
M. L\'opez-Corredoira, A. Vazdekis, C.M. Guti\'errez and N. Castro-Rodr\'iguez, {\it Stellar content of extremely red quiescent galaxies at $z>2$}, Astron. Astrophys. {\bf 600} (2017) A91 [arXiv:1702.00380]

\bibitem{LopezCorredoira2018}
M. L\'opez-Corredoira and A. Vazdekis, {\it Impact of young stellar components on quiescent galaxies: deconstructing cosmic chronometers}, [arXiv:1802.09473]

%%%%%%%%%%%%%%%%%%%%%%%%%%  Inflation %%%%%%%%%%%%%%%%%%%%%%%%%%%%%%%%

\bibitem{linde}
A. D. Linde, ``Chaotic Inflation'', Phys.Lett. {\bf B 129}(177-181) (1983).

\bibitem{Kolb}
E. W. Kolb, M. S. Turner, ``The Early Universe'', Front.Phys. {\bf 69}(1-547) (1990).

\bibitem{mrg1}
G.~J.~Mathews, M.~R.~Gangopadhyay, K.~Ichiki and T.~Kajino,
{\it Possible Evidence for Planck-Scale Resonant Particle Production during Inflation from the CMB Power Spectrum,}
Phys. Rev. D \textbf{92}, no.12, 123519 (2015)
[arXiv:1504.06913 [astro-ph.CO]].

\bibitem{mar1}
M.~Bastero-Gil, S.~Bhattacharya, K.~Dutta and M.~R.~Gangopadhyay,
{\it Constraining Warm Inflation with CMB data},
JCAP \textbf{02}, 054 (2018), arXiv:1710.10008 [astro-ph.CO].

\bibitem{sharma}
M.~R.~Gangopadhyay, S.~Myrzakul, M.~Sami and M.~K.~Sharma,
{\it Paradigm of warm quintessential inflation and production of relic gravity waves}, Phys. Rev. D \textbf{103}, no.4, 043505 (2021), arXiv:2011.09155 [astro-ph.CO].

\bibitem{Bhattacharya}
S.~Bhattacharya and M.~R.~Gangopadhyay,
{\it Study in the noncanonical domain of Goldstone inflation},
Phys. Rev. D \textbf{101}, no.2, 023509 (2020)[arXiv:1812.08141 [astro-ph.CO]].

\bibitem{Bhattacharya:2020zap1}
S.~Bhattacharya, S.~Das, K.~Dutta, M.~R.~Gangopadhyay, R.~Mahanta and A.~Maharana, {\it Nonthermal hot dark matter from inflaton or moduli decay: Momentum distribution and relaxation of the cosmological mass bound},
\ Phys.\ Rev. D \textbf{103}, no.6, 063503 (2021), arXiv:2009.05987 [astro-ph.CO].

\bibitem{starobinsky}
A. A. Starobinsky, ``A New Type of Isotropic Cosmological Models Without Singularity'', Phys.Lett. {\bf B 91}(99-102) (1980). 

%%%%%%%%%%%%%%%%%%%%%%%%%% Scalar field %%%%%%%%%%%%%%%%%%%%%%%%%%%%%%%

\bibitem{Ratra} 
P.~J.~E.~Peebles and B.~Ratra, ``The Cosmological constant and dark energy,''
Rev.\ Mod.\ Phys.\ \textbf{75}, 559 (2003), astro-ph/0207347.

\bibitem{Wetterich} 
C.~Wetterich, ``Cosmology and the Fate of Dilatation Symmetry,''Nucl.\ Phys.\ B \textbf{302}, 668 (1988). 

\bibitem{Ratra2} 
B.~Ratra and P.~J.~E.~Peebles, ``Cosmological Consequences of a Rolling Homogeneous Scalar Field,''Phys.\ Rev.\ D \textbf{37}, 3406 (1988). 

\bibitem{Steinhardt} 
R.~R.~Caldwell, R.~Dave and P.~J.~Steinhardt, ``Cosmological imprint of an energy component with general equation of state,''Phys.\ Rev.\ Lett.\ \textbf{80}, 1582 (1998), astro-ph/9708069.

%%%%%%%%%%%%%%%%%%%%%%%%%%%  Scalar-tensor  %%%%%%%%%%%%%%%%%%%%%%%

\bibitem{ijmpds} 
M. Sami, ``Modified Theories of Gravity and Constraints Imposed by Recent GW Observations'', IJMPD {\bf 28}(Special Issue), Number {\bf 5} (2019).


\bibitem{SAMI-NOJ} 
K. Bamba, Md. Wali Hossain, R. Myrzakulov, S. Nojiri, M.
Sami, Phys. Rev. {\bf D} \textbf{89}, 083518 (2014), arXiv:1309.6413 [hep-th]; M. Sami, R. Myrzakulov, Int.J.Mod.Phys.D 25 (2016) 12, 1630031.

%%%%%%%%%%%%%%%%%%%%%%%%%%%% SN1a %%%%%%%%%%%%%%%%%%%%%%%%%%%%

\bibitem{Riess} 
A.~G.~Riess \textit{et al.} [Supernova Search Team], 
``Observational evidence from supernovae for an accelerating universe and a cosmological constant'',
Astron.\ J.\ \textbf{116}, 1009 (1998), astro-ph/9805201. 

\bibitem{Perlmutter} 
S.~Perlmutter \textit{et al.} [Supernova Cosmology Project Collaboration], ``Measurements of Omega and Lambda from 42 high redshift supernovae'', Astrophys.\ J.\ \textbf{517}, 565 (1999), astro-ph/9812133. 

\bibitem{jla} 
M. Betoule et al., Improved cosmological constraints from a joint analysis of the SDSS-II and SNLS supernova samples, Astron. Astrophys. 568 (2014) A22, arXiv:1401.4064.

\bibitem{pantheon}
D.M. Scolnic et al., The Complete Light-curve Sample of Spectroscopically Confirmed Type Ia Supernovae from Pan-STARRS1 and Cosmological Constraints from The Combined Pantheon Sample, arXiv:1710.00845.

%%%%%%%%%%%%%%%%%%%%%%%%%%%% Low-z Observations %%%%%%%%%%%%%%%%%%%%%%%%%%%%

\bibitem{Seljak} 
U.~Seljak \textit{et al.} [SDSS Collaboration], 
``Cosmological parameter analysis including SDSS Ly-alpha forest and galaxy bias: Constraints on the primordial spectrum of fluctuations, neutrino mass, and dark energy'', Phys.\ Rev.\ {\bf D} \textbf{71}, 103515 (2005), astro-ph/0407372. 

\bibitem{BasilakosH0}
D. Fern\'andez-Arenas {\it et al.}, {\it An independent determination of the local Hubble constant}, Mont. Not. Roy. Astron. Soc. {\bf 474} (2018) 1, arXiv:1710.05951.

\bibitem{reiss2019}
Adam G. Riess {\it et al}, {\it Large Magellanic Cloud Cepheid Standards Provide a 1\% Foundation for the Determination of the Hubble Constant and Stronger Evidence for Physics beyond $\Lambda$CDM}, ApJ \textbf{876}, 85(2019), arXiv:1903.07603.

\bibitem{reiss2021}
Adam G. Riess {\it et al.}, {\it Cosmic Distances Calibrated to 1\% Precision with Gaia EDR3 Parallaxes and Hubble Space Telescope Photometry of 75 Milky Way Cepheids Confirm Tension with $\Lambda$CDM}, ApJL \textbf{908}, L6(2021).

\bibitem{SBF}
John P. Blakeslee, Joseph B. Jensen, Chung-Pei Ma, Peter A. Milne, Jenny E. Greene, {\it The Hubble Constant from Infrared Surface Brightness Fluctuation Distances}, ApJ \textbf{911}, 65(2021), arXiv:2101.02221.

\bibitem{trgb}
Wendy L. Freedman {\it et al.}, {\it The Carnegie-Chicago Hubble Program. VIII. An Independent Determination of the Hubble Constant Based on the Tip of the Red Giant Branch}, ApJ \textbf{882}, 34(2019), arXiv:1907.05922.

\bibitem{birrer}
S. Birrer {\it et al.}, {\it H0LiCOW - IX. Cosmographic analysis of the doubly imaged quasar SDSS 1206+4332 and a new measurement of the Hubble constant}, Mon. Not. R. Astron. Soc. \textbf{484}, 4726–4753 (2019), arXiv:1809.01274.

\bibitem{chen}
Geoff C.-F. Chen {\it et al.}, {\it A SHARP view of H0LiCOW: H0 from three time-delay gravitational lens systems with adaptive optics imaging}, Mon. Not. R. Astron. Soc. \textbf{490}, 2(2019), arXiv:1907.02533.

\bibitem{wong}
Kenneth C. Wong {\it et al.}, {\it H0LiCOW XIII. A 2.4\% measurement of H0 from lensed quasars: 5.3$\sigma$ tension between early and late-Universe probes}, Mon. Not. R. Astron. Soc. \textbf{498}, 1(2020), arXiv:1907.04869.

\bibitem{huang}
Caroline D. Huang {\it et al.}, {\it Hubble Space Telescope Observations of Mira Variables in the Type Ia Supernova Host NGC 1559: An Alternative Candle to Measure the Hubble Constant}, ApJ \textbf{889}, 5(2020), arXiv:1908.10883. 

\bibitem{snII}
T de Jaeger, B E Stahl, W Zheng, A V Filippenko, A G Riess, L Galbany, {\it A measurement of the Hubble constant from Type II supernovae}, Mon. Not. R. Astron. Soc. \textbf{496}, 3(2020), arXiv:2006.03412.

\bibitem{TFR}
James Schombert, Stacy McGaugh, Federico Lelli, {\it Using The Baryonic Tully-Fisher Relation to Measure H0}, AJ \textbf{160}, 71(2020), arXiv:2006.08615.

\bibitem{hotens}
E.~Di Valentino, O.~Mena, S.~Pan, L.~Visinelli, W.~Yang, A.~Melchiorri, D.~F.~Mota, A.~G.~Riess and J.~Silk,
{\it In the Realm of the Hubble tension $-$ a Review of Solutions},  arXiv:2103.01183 [astro-ph.CO].

\bibitem{Efstathiou2014}
G. Efstathiou, {\it $H_0$ revisited}, Mont. Not. Roy. Astron. Soc. {\bf 440} (2014) 1138, arXiv:1311.3461.

\bibitem{Hub19}
Rafael C. Nunes, {\it Structure formation in f(T) gravity and a solution for H0 tension}, JCAP \textbf{05}, 052(2018), arXiv:1802.02281.

\bibitem{RiessH02018}
A.G. Riess {\it et al.}, {\it New Parallaxes of Galactic Cepheids from Spatially Scanning the Hubble Space Telescope: Implications for the Hubble Constant}, arXiv:1801.01120.


\bibitem{Riess2011}
A.G. Riess {\it et al.}, {\it A 3\% Solution: Determination of the Hubble Constant with the Hubble Space Telescope and Wide Field Camera 3}, Astrophys. J. {\bf 730} (2011) 119 [{\it Erratum ibid} {\bf 732} (2011) 129], arXiv:1103.2976.

\bibitem{RiessH02016}
A.G. Riess {\it et al.}, {\it A $2.4\%$ Determination of the Local Value of the Hubble Constant}, Astrophys. J. {\bf 826} (2016) 56, arXiv:1604.01424.

\bibitem{RiessH02017}
R.I. Anderson and A.G. Riess, {\it On Cepheid distance scale bias due to stellar companions and cluster populations}, arXiv:1712.01065.

\bibitem{Melchiorri2016}
E.D. Valentino, A. Melchiorri and J. Silk, {\it Reconciling Planck with the local value of $H_0$ in extended parameter space}, Phys. Lett. B{\bf 761} (2016) 242 [arXiv:1606.00634]

\bibitem{Jarahetal2017}
Jarah Evslin, Anjan A Sen, Ruchika, {\it The Price of Shifting the Hubble Constant}, Phys. Rev. D \textbf{97}, 103511 (2018) [	arXiv:1711.01051 [astro-ph.CO]]

\bibitem{Feeney2018}
S.M. Feeney {\it et al.}, {\it Prospects for resolving the Hubble constant tension with standard sirens}, arXiv:1802.03404.

\bibitem{holicow}
K. C. Wong {\it et al.}, Mon. Not. R. Astron. Soc. {\bf 498} 1420–3 (2020).

\bibitem{megamaser}
D. W. Pesce {\it et al.}, Astrophys. J.{\bf 891} L1 (2020).

\bibitem{Reidetal2013}
M. J. Reid, J. A. Braatz, J. J. Condon, K. Y. Lo, C. Y. Kuo, C. M. V. Impellizzeri and C. Henkel, “The Megamaser Cosmology Project: IV. A Direct Measurement of the Hubble Constant from UGC 3789,” Astrophys. J. 767 (2013) 154 doi:10.1088/0004- 637X/767/2/154 [arXiv:1207.7292 [astro-ph.CO]].

\bibitem{Kuoetal2013}
C. Kuo, J. A. Braatz, M. J. Reid, F. K. Y. Lo, J. J. Condon, C. M. V. Impellizzeri and C. Henkel, “The Megamaser Cosmology Project. V. An Angular Diameter Dis- tance to NGC 6264 at 140 Mpc,” Astrophys. J. 767 (2013) 155 doi:10.1088/0004- 637X/767/2/155 [arXiv:1207.7273 [astro-ph.CO]].

\bibitem{Gaoetal2016}
F. Gao et al., “The Megamaser Cosmology Project VIII. A Geometric Distance to NGC 5765b,” Astrophys. J. 817 (2016) no.2, 128 doi:10.3847/0004-637X/817/2/128 [arXiv:1511.08311 [astro-ph.GA]].

\bibitem{kam}
T. Karwal and M. Kamionkowski, Phys. Rev. {\bf D} {\bf 94} 103523 (2016).

\bibitem{kam2}
V. Poulin, T. L. Smith, T. Karwal, M. Kamionkowski, Phys. Rev. Lett. {\bf 122} 221301 (2019).

\bibitem{trodden}
J. Sakstein and M. Trodden, Phys. Rev. Lett. {\bf 124} 161301 (2020).

\bibitem{baren}
G. Barenboim {\it et al.}, Eur. Phys. J. {\bf C} {\bf 77} 590 (2017).
 
\bibitem{pogosian}
K. Jedamzik, and L. Pogosian, Phys. Rev. Lett. {\bf 125} 181302 (2020).

\bibitem{pogosian2}
K. Jedamzik, L. Pogosian, and G. B. Zhao, Commun.Phys. {\bf 4} 123 (2021).
 
\bibitem{perivola}
G. Alestas and L. Perivolaropoulos, Mon. Not. R. Astron.Soc. {\bf 504} 3956 (2021).
 
\bibitem{Benevento} 
G. Benevento, W. Hu, M. Raveri, Phys. Rev. {\bf D} {\bf 101} 103517 (2020).
 
%%%%%%%%%%%%%%%%%%%%%%%%%%% Modified Gravity %%%%%%%%%%%%%%%%%%%%%%%%%%%%

\bibitem{antonio}
A. D. Felice, S. Tsujikawa, ''f(R) theories'', Living Rev. Rel. {\bf 13}(3) (2010), arXiv:1002.4928[gr-qc].

\bibitem{derham}
C. d. Rham, G. Gabadadze, and A. J. Tolley, ``Resummation of Massive Gravity'', Phys. Rev. Lett. {\bf 106}(231101) (2011), arXiv:1011.1232[hep-th]. 

\bibitem{massive2}
K. Hinterbichler, ``Theoretical Aspects of Massive Gravity'', Rev. Mod. Phys. {\bf 84}(671-710) (2012), arXiv:1105.3735[hep-th].

\bibitem{massive3}
S. Deser, K. Izumi, Y. C. Ong, A. Waldron, ``Problems of Massive Gravities'', Mod. Phys. Lett. {\bf A 30}(1540006) (2015), arXiv:1410.2289[hep-th].

\bibitem{massive4}
S.  Deser,  A.  Waldron, ``Acausality  of  Massive  Gravity'',  Phys.  Rev.  Lett. {\bf 110}(111101) (2013), arXiv:1212.5835v3 [hep-th].

\bibitem{KHOURY2016} 
L.~Berezhiani, J.~Khoury and J.~Wang, 
``Universe without dark energy: Cosmic acceleration from dark matter-baryon interactions,''Phys.\ Rev.\ D \textbf{95}, no. 12, 123530 (2017), arXiv:1612.00453 [hep-th]. 

\bibitem{shibesh}
A. Agarwal , R. Myrzakulov , S.K.J. Pacif, M. Sami , Anzhong Wang,
``Cosmic acceleration sourced by modification of gravity without extra degrees of freedom'', Int.J.Geom.Meth.Mod.Phys. {\bf 16} (2019) no.08, 1950128,
 arXiv:1709.02133 [gr-qc]; components: Analysis and diagnostics
Abhineet Agarwal, R. Myrzakulov, S.K. J. Pacif, M. Shahalam, Int. J. Mod. Phys. D {\bf}28 (2019) no.06, 1950083. 

\bibitem{jwang} 
J.~Wang, L.~Hui and J.~Khoury, ``No-Go Theorems for Generalized Chameleon Field Theories,''Phys.\ Rev.\ Lett.\ \textbf{109}, 241301 (2012) [arXiv:1208.4612 [astro-ph.CO]]. 

\bibitem{kBamba} 
K.~Bamba, R.~Gannouji, M.~Kamijo, S.~Nojiri and M.~Sami, 
``Spontaneous symmetry breaking in cosmos: The hybrid symmetron as a dark energy switching device,'' JCAP \textbf{1307}, 017 (2013) [arXiv:1211.2289 [hep-th]]. 

\bibitem{LPC1} 
A.~Upadhye, W.~Hu and J.~Khoury, ``Quantum Stability of Chameleon Field Theories,''Phys.\ Rev.\ Lett.\ \textbf{109}, 041301 (2012), arXiv:1204.3906 [hep-ph]. 

\bibitem{vainsh1}
A. Joyce, B. Jain, J. Khoury, and M. Trodden, ``Beyond the Cosmological Standard Model'', Phys. Rept. {\bf 568}(1) (2015), arXiv:1407.0059[astro-ph.CO].

\bibitem{vainsh2}
E. Babichev, C. Deffayet, ``An introduction to the Vainshtein mechanism'', Class. Quant. Grav. {\bf 30}(184001) (2013), arXiv:1304.7240[gr-qc].

\bibitem{chamel0} 
D.~F.~Mota and J.~D.~Barrow, 
``Varying alpha in a more realistic Universe,''
Phys.\ Lett.\ B \textbf{581}, 141 (2004) [astro-ph/0306047].

\bibitem{chamel1} 
J.~Khoury and A.~Weltman, ``Chameleon fields: Awaiting surprises for tests of gravity in space,''Phys.\ Rev.\ Lett.\ \textbf{93}, 171104 (2004), 
astro-ph/0309300. 

\bibitem{chamel2} 
P.~Brax, C.~van de Bruck, A.~C.~Davis, J.~Khoury and
A.~Weltman, ``Detecting dark energy in orbit - The Cosmological chameleon,''
Phys.\ Rev.\ D \textbf{70}, 123518 (2004), astro-ph/0408415. 

\bibitem{EXTENDED} 
S. Capozziello, R. D'Agostino, O. Luongo, (2019), arXiv:1904.01427 [gr-qc].

\bibitem{CAPOZZI} 
S. Capozziello, S. Nojiri, S. D. Odintsov, A. Troisi,
Phys. Lett. B\textbf{\ 639}, 135 (2006), arXiv:astro-ph/0604431v3.

\bibitem{sponts}
K. Bamba1,2, R. Gannouji, M. Kamijo, S. Nojiri1, and M. Sami,JCAP 07(2013) 017.

\bibitem{ballardini1}
Mario Ballardini, Matteo Braglia, Fabio Finelli, Daniela Paoletti,
Alexei A. Starobinsky, Caterina Umiltà, {\it Scalar-tensor theories of gravity, neutrino physics, and the H0 tension}, JCAP \textbf{10}, 044(2020), arXiv:2004.14349.

\bibitem{ballardini3}
C. Umiltà, M. Ballardini, F. Finelli, D. Paoletti,
{\it CMB and BAO constraints for an induced gravity dark energy model with a quartic potential}, JCAP \textbf{08}, 017(2015), [arXiv:1507.00718]

\bibitem{ballardini2}
Matteo Braglia, Mario Ballardini, Fabio Finelli, Kazuya Koyama, {\it Early modified gravity in light of the H0 tension and LSS data}, Phys.Rev. {\bf D} \textbf{103}, no. 4, 043528(2021), arXiv:2011.12934.

\bibitem{ballardini4}
Mario Ballardini, Fabio Finelli, Caterina Umiltà, Daniela Paoletti,
{\it Cosmological constraints on induced gravity dark energy models}, JCAP \textbf{05}, 067(2016). [arXiv:1601.03387]

%%%%%%%%%%%%%%%%%%%%%%%%%%%%%%%%%%%%%%%%%%%%%%%%%%%%%%%%%%%%%%%%%%%%%%%%%%

\bibitem{shah} 
S. A. Adil, M. R. Gangopadhyay , M. Sami, M. K. Sharma, Phys. 
Rev. {\bf D} 104 (2021) 10, 10353, arXiv:2106.03093[astro-ph.CO].

\bibitem{Planck18}
N.~Aghanim \textit{et al.} [Planck], {\it Planck 2018 results. VI. Cosmological parameters}, Astron. Astrophys. \textbf{641}, A6 (2020), arXiv:1807.06209 [astro-ph.CO].
%%%%%%%%%%%%%%%%%%%%%%%%%% Hubble Tension %%%%%%%%%%%%%%%%%%%%%%%%%%%%%%%%



\bibitem{Hub2}
R. R. Caldwell ,M. Doran ,C. M. Mueller ,G. Schafer  and C. Wetterich, 
{\it Early Quintessence in Light of WMAP}  ,Astrophys. J. Lett. \textbf{591},
L75–L78(2003), astro-ph/0302505.

%
\bibitem{Hub3}
T. Karwal  and M. Kamionkowski ,{\it Early dark energy, the Hubble-parameter tension, and the string axiverse} Phys. Rev. {\bf D} \textbf{94}, 103523(2016), arXiv:1608.01309.

\bibitem{Hub4}
V. Pettorino ,L. Amendola and C. Wetterich, {\it How early is early dark energy?}, Phys. Rev. {\bf D} \textbf{87} 083009(2013), arXiv:1301.5279.

\bibitem{Hub5}
V. Poulin ,T. L. Smith,T. Karwal T and M. Kamionkowski, {\it Early Dark Energy Can Resolve The Hubble Tension} , Phys. Rev. Lett. \textbf{122}, 221301(2019),
arXiv:1811.04083.

\bibitem{Hub6}
M. Kamionkowski ,J. Pradler and D G E Walker, {\it Dark energy from the string axiverse}, Phys. Rev. Lett. \textbf{113}, 251302(2014), arXiv:1409.0549.


\bibitem{Hub7}
V. Poulin, T. L. Smith, D. Grin, T. Karwal and M. Kamionkowski, {\it Cosmological implications of ultra-light axion-like fields}, Phys. Rev. D \textbf{98}, 083525(2018), arXiv:1806.10608.

\bibitem{Hub8}
A. Banerjee , H. Cai, L. Heisenberg, E. O. Colgáin, M. M. Sheikh-Jabbari and T. Yang T, {\it Hubble Sinks In The Low-Redshift Swampland}, arXiv:2006.00244.

\bibitem{PLB2017}
J. Sol\`a, A. G\'omez-Valent and J. de Cruz P\'erez, {\it The $H_0$ tension in light of vacuum dynamics in the Universe}, Phys. Lett. B{\bf 774} (2017) 317, arXiv:1705.06723.

\bibitem{Hub9}
 U. Alam, S. Bag and V. Sahni, {\it Constraining the Cosmology of the Phantom Brane using Distance Measures} ,Phys. Rev. {\bf D} \textbf{95}, 023524(2017), arXiv:1605.04707.

\bibitem{Hub10}
E. Di Valentino ,V. E. Linder and A. Melchiorri, {\it A Vacuum Phase Transition Solves H0 Tension} , Phys. Rev. D \textbf{97}, 043528(2018), arXiv:1710.02153.

\bibitem{Hub11}
K. L. Pandey , T.  Karwal and S. Das, {\it Alleviating the $H_0$ and $\sigma_8$ anomalies with a decaying dark matter model}, JCAP \textbf{07}, 026(2020), arXiv:1902.10636.

\bibitem{Hub12}
Antareep Gogoi, Ravi Kumar Sharma, Prolay Chanda, Subinoy Das,
{\it Early mass varying neutrino dark energy: Nugget formation and Hubble anomaly}, arXiv:2005.11889.

\bibitem{Hub17}
S. Kumar, R. C. Nunes, S. K. Yadav, {\it Dark sector interaction: a remedy of the tensions between CMB and LSS data}, Eur. Phys. J. {\bf C} \textbf{79}, 576(2019), arXiv:1903.04865.

\bibitem{Melchiorri2017b}
E.D. Valentino, A. Melchiorri and O. Mena, {\it Can interacting dark energy solve the $H_0$ tension?}, Phys. Rev. D{\bf 96} (2017) 043503, arXiv:1704.08342.

\bibitem{Hub16}
Maria Giovanna Dainotti, Biagio De Simone, Tiziano Schiavone, Giovanni Montani, Enrico Rinaldi, Gaetano Lambiase, {\it On the Hubble constant tension in the SNe Ia Pantheon sample}, ApJ \textbf{912}, 150(2021), arXiv:2103.02117.

\bibitem{TammannReindl2013}
G.A. Tammann and B. Reindl, {\it The luminosity of supernovae of type Ia from TRGB distances and the value of $H_0$}, Astron. Astrophys. {\bf 549} (2013) A136, arXiv:1208.5054.

\bibitem{JangLee2017}
S. Jang and M.G. Lee, {\it The Tip of the Red Giant Branch Distances to Type Ia Supernova Host Galaxies. V. NGC 3021, NGC 3370, and NGC 1309 and the Value of the Hubble Constant}, Astrophys. J. {\bf 836} (2017) 74, arXiv:1702.01118.

\bibitem{Blake:2011en} 
C.~Blake, E.~Kazin, F.~Beutler, T.~Davis,
D.~Parkinson, S.~Brough, M.~Colless and C.~Contreras \textit{et al.}, 
``The WiggleZ Dark Energy Survey: mapping the distance-redshift relation
with baryon acoustic oscillations,''Mon.\ Not.\ Roy.\ Astron.\ Soc.\ \textbf{418}, 1707 (2011), arXiv:1108.2635[astro-ph.CO].

\bibitem{Jarosik:2010iu} 
N.~Jarosik, C.~L.~Bennett, J.~Dunkley, B.~Gold,
M.~R.~Greason, M.~Halpern, R.~S.~Hill and G.~Hinshaw \textit{et al.}, 
``Seven-Year Wilkinson Microwave Anisotropy Probe (WMAP) Observations:
sky Maps, Systematic Errors, and Basic Results,''
Astrophys.\ J.\ Suppl.\ \textbf{192}, 14 (2011), arXiv:1001.4744
[astro-ph.CO]. 

\bibitem{Freedman2012}
W.L. Freedman {\it et al.}, {\it Carnegie Hubble Program: A Mid-Infrared Calibration of the Hubble Constant}, Astrophys. J. {\bf 758} (2012) 24, arXiv:1208.3281.

\bibitem{Casertano2017}
S. Casertano, A.G. Riess, B. Bucciarelli and M.G. Lattanzi, {\it A test of Gaia Data Release 1 parallaxes: implications for the local distance scale}, Astron. Astrophys. {\bf599} (2017) A67, arXiv:1609.05175.

\bibitem{Cardona2017} 
W. Cardona, M. Kunz and V. Pettorino, {\it Determining $H_0$ with Bayesian hyper-parameters}, J. Cosmol. Astropart. Phys. {\bf 1703} (2017) 056, arXiv:1611.06088.

\bibitem{Zhang2017}
B.R. Zhang {\it et al.}, \textit{A blinded determination of $H_0$ from low-redshift Type Ia supernovae}, Mon. Not. Roy. Astron. Soc. {\bf 471} (2017) 2254, arXiv:1706.07573.

\bibitem{Feeney2017}
S.M. Feeney, D.J. Mortlock and N. Dalmasso, {\it Clarifying the Hubble constant tension with a Bayesian hierarchical model of the local distance ladder}, Mon. Not. Roy. Astron. Soc. {\bf476} (2018) 3861, arXiv:1707.00007.

\bibitem{SuhailDhawan2018}
S. Dhawan, S.W. Jha and B. Leibundgut, {\it Measuring the Hubble constant with Type Ia supernovae as near-infrared standard candles}, Astron. Astrophys. {\bf 609} (2018) A72, arXiv:1707.00715. 

\bibitem{Bonvin2017}
V. Bonvin {\it et al.}, {\it H0LiCOW - V. New COSMOGRAIL time delays of HE $0435-1223$: $H_0$ to $3.8$ per cent precision from strong lensing}, Mon. Not. Roy. Astron. Soc. {\bf 465} (2017) 4914, arXiv:1607.01790.

\bibitem{WangMeng2017}
D. Wang and X.-H. Meng, {\it Determining $H_0$ with the latest HII galaxy measurements}, Astrophys. J. {\bf 843} (2017) 100, arXiv:1612.09023.

\bibitem{Lin2017}
W. Lin and M. Ishak, \textit{Cosmological discordances II: Hubble constant, Planck and large-scale-structure data sets}, Phys. Rev. D{\bf96} (2017) 083532, arXiv:1708.09813.


\bibitem{Aylor2017}
A. Aylor {\it et al.}, {\it A comparison of cosmological parameters determined from CMB temperature power spectra from the South Pole telescope and the Planck Satellite}, Astrophys. J. {\bf 850} (2017) 101, arXiv:1706.10286

\bibitem{Addison2017}
G.E. Addison {\it et al.}, {\it Elucidating $\Lambda$CDM: impact of baryon acoustic oscillation measurements on the Hubble constant discrepancy}, Astrophys. J. {\bf 853} (2018) 119, arXiv:1707.06547.

\bibitem{Follin2017}
B. Follin and L. Knox, {\it Insensitivity of The Distance Ladder Hubble Constant Determination to Cepheid Calibration Modeling Choices}, accepted for publication in Mont. Not. Roy. Astron. Soc., arXiv:1707.01175.

\bibitem{Marra2013}
V. Marra, L. Amendola, I. Sawicky and W. Valkenburg, {\it Cosmic variance and the measurement of the local Hubble parameter}, Phys. Rev. Lett. {\bf 110} (2013) 241305, arXiv:1303.3121.

\bibitem{Wojtak2014}
R. Wojtak {\it et al.}, {\it Cosmic variance of the local Hubble flow in large-scale cosmological simulations}, Mont. Not. Roy. Astron. Soc. {\bf 438} (2014) 1805, arXiv:1312.0276.


\bibitem{Melchiorri2017}
E.D. Valentino, A. Melchiorri, E.V. Linder and J. Silk, {\it Constraining dark energy dynamics in extended parameter space}, Phys. Rev. {\bf D} {\bf 96} (2017) 023523, arXiv:1704.00762.

\bibitem{VerdeRiessH0}
J.L. Bernal, L. Verde and A.G. Riess, {\it The trouble with $H_0$}, J. Cosmol. Astropart. Phys. {\bf 1610} (2016) 019, arXiv:1607.05617.

\bibitem{MortsellDhawan2018}
E. M\"ortsell and S. Dhawan, {\it Does the Hubble constant tension call for new physics?}, arXiv:1801.07260.


\bibitem{Wyman2014}
M. Wyman, D.H. Rudd, R.A.Vanderveld and W. Hu, {\it Neutrinos Help Reconcile Planck Measurements with the Local Universe}, Phys. Rev. Lett. {\bf 112} (2014) 051302, arXiv:1307.7715. 

\bibitem{Poulin2018}
V. Poulin, K.K. Boddy, S. Bird and M. Kamionkowski, {\it The implications of an extended dark energy cosmology with massive neutrinos for cosmological tensions}, arXiv:1803.02474.

\bibitem{EPL2017}
A. G\'omez-Valent and J. Sol\`a, {\it Relaxing the $\sigma_8$-tension through running vacuum in the Universe}, Europhys. Lett. {\bf 120} (2017) 39001, arXiv:1711.00692.

\bibitem{MNRAS2018}                                 
A. G\'omez-Valent and J. Sol\`a, {\it Density perturbations for running vacuum: a successful approach to structure formation and to the $\sigma_8$-tension}, arXiv:1801.08501.

\bibitem{ApJL2015}
J. Sol\`a, A. G\'omez-Valent and J. de Cruz P\'erez, {\it Hints of dynamical vacuum energy in the expanding Universe}, Astrophys. J. {\bf811} (2015) L14, arXiv:1506.05793.

\bibitem{ApJ2017}
J. Sol\`a, A. G\'omez-Valent and J. de Cruz P\'erez, {\it First evidence of running cosmic vacuum: challenging the concordance model}, Astrophys. J. {\bf836} (2017) 43, arXiv:1602.02103.

\bibitem{EPL2018x3}
J. Sol\`a, J. de Cruz P\'erez and A. G\'omez-Valent, {\it Dynamical dark energy versus $\Lambda=const.$ in light of observations}, Europhys. Lett. {\bf121} (2018) 39001, arXiv:1606.00450.

\bibitem{ChenRatra2011}
G. Chen and B. Ratra, {\it Median statistics and the Hubble constant}, Publ. Astron. Soc. Pac. {\bf 123} (2011) 1127, arXiv:1105.5206.


\bibitem{GWH0second}
C. Guidorzi {\it et al.}, {\it Improved constraints on $H_0$ from a combined analysis of gravitational-wave and electromagnetic emission from $GW170817$}, Astrophys. J. {\bf 851} (2017) L36, arXiv:1710.06426.

%%%%%%%%%%%%%%%%%%%%%%%%%%%%%%%%%%%%%%%%%%%%%%%%%%%%%%%%%%%%%%%%%%%%%%%%%%%%

\bibitem{riess}
A.G. Riess et al., Type Ia Supernova distances at redshift > 1.5 from the Hubble Space Telescope Multy-Cycle Treasury programs: the early expansion rate, Astrophys. J. 853 (2018) 126 [arXiv:1710.00844]

\bibitem{amendola}
Gomez-Valent A. and Amendola L., 2018, JCAP, 1804, 051

\bibitem{Morescoetal2012}
M. Moresco et al., Improved constraints on the expansion rate
of the Universe up to z $\sim$ 1.1 from the spectroscopic evolution
of cosmic chronometers, JCAP 08, 006 (2012). 1201.3609.

\bibitem{Morescoetal2016}
M. Moresco, L. Pozzetti, A. Cimatti, R. Jimenez, C. Maraston, L. Verde, D. Thomas, A. Citro, R. Tojeiro, and
D. Wilkinson, {\it A 6\% measurement of the Hubble parameter at $z \sim 0.45$: Direct evidence of the epoch of cosmic re-acceleration}, {JCAP \textbf{05}, 014 (2016)}. arXiv : 1601.01701

\bibitem{Moresco2015}
M. Moresco, {\it Raising the bar: new constraints on the Hubble parameter with cosmic chronometers at z$\sim$ 2}, {Mon. Not. R. Astron. Soc. Lett. \textbf{450}, L16 (2015)}. arXiv : 1503.01116

\bibitem{Lewis}
Antony Lewis, {\it GetDist: a Python package for analysing Monte Carlo samples}, [	arXiv:1910.13970 [astro-ph.IM]]

\bibitem{riess2018}
Adam G. Riess {\it et. al.} , {\it Milky Way Cepheid Standards for Measuring Cosmic Distances and Application to Gaia DR2: Implications for the Hubble Constant}, ApJ \textbf{861}, 126(2018). [	arXiv:1804.10655 [astro-ph.CO]].

\bibitem{Ly1a}
Timothée Delubac {\it et. al.}, {\it Baryon Acoustic Oscillations in the Ly$\alpha$ forest of BOSS DR11 quasars}, A \& A \textbf{574}, A59 (2015) [arXiv:1404.1801 [astro-ph.CO]].
 


\end{thebibliography}
\end{document}